\documentclass[%
 reprint,
superscriptaddress,
frontmatterverbose, 
showpacs,preprintnumbers,
 amsmath,amssymb,
 aps,
]{revtex4-1}
\usepackage{xcolor}
\usepackage{graphicx}
\usepackage{dcolumn}
\usepackage{bm}

\begin{document}

\title{Development and Operation of a \textbf{$Pr_2Fe_{14}B$} Based Cryogenic Permanent Magnet Undulator for a High Spatial Resolution X-ray Beamline}

\author{C. Benabderrahmane}

\affiliation{Synchrotron SOLEIL, L'Orme des Merisiers, Bat. A, Saint-Aubin, Gif-sur-Yvette 91192, FRANCE.}
\altaffiliation[Also at ]{ESRF 71 Avenue des martyrs, Grenoble 38000, FRANCE.}

\author{M. Vall\'eau}
\affiliation{Synchrotron SOLEIL, L'Orme des Merisiers, Bat. A, Saint-Aubin, Gif-sur-Yvette 91192, FRANCE.}

\author{A. Ghaith}
\email{amin.ghaith@synchrotron-soleil.fr}
\affiliation{Synchrotron SOLEIL, L'Orme des Merisiers, Bat. A, Saint-Aubin, Gif-sur-Yvette 91192, FRANCE.}

\author{P. Berteaud}
\affiliation{Synchrotron SOLEIL, L'Orme des Merisiers, Bat. A, Saint-Aubin, Gif-sur-Yvette 91192, FRANCE.}

\author{L. Chapuis}
\affiliation{Synchrotron SOLEIL, L'Orme des Merisiers, Bat. A, Saint-Aubin, Gif-sur-Yvette 91192, FRANCE.}

\author{F. Marteau}
\affiliation{Synchrotron SOLEIL, L'Orme des Merisiers, Bat. A, Saint-Aubin, Gif-sur-Yvette 91192, FRANCE.}

\author{F. Briquez}
\affiliation{Synchrotron SOLEIL, L'Orme des Merisiers, Bat. A, Saint-Aubin, Gif-sur-Yvette 91192, FRANCE.}

\author{O. Marcouill\'e}
\affiliation{Synchrotron SOLEIL, L'Orme des Merisiers, Bat. A, Saint-Aubin, Gif-sur-Yvette 91192, FRANCE.}

\author{J.-L. Marlats}
\affiliation{Synchrotron SOLEIL, L'Orme des Merisiers, Bat. A, Saint-Aubin, Gif-sur-Yvette 91192, FRANCE.}

\author{K. Tavakoli}
\affiliation{Synchrotron SOLEIL, L'Orme des Merisiers, Bat. A, Saint-Aubin, Gif-sur-Yvette 91192, FRANCE.}

\author{A. Mary}
\affiliation{Synchrotron SOLEIL, L'Orme des Merisiers, Bat. A, Saint-Aubin, Gif-sur-Yvette 91192, FRANCE.}

\author{D. Zerbib}
\affiliation{Synchrotron SOLEIL, L'Orme des Merisiers, Bat. A, Saint-Aubin, Gif-sur-Yvette 91192, FRANCE.}

\author{A. Lestrade}
\affiliation{Synchrotron SOLEIL, L'Orme des Merisiers, Bat. A, Saint-Aubin, Gif-sur-Yvette 91192, FRANCE.}

\author{M. Louvet}
\affiliation{Synchrotron SOLEIL, L'Orme des Merisiers, Bat. A, Saint-Aubin, Gif-sur-Yvette 91192, FRANCE.}

\author{P. Brunelle}
\affiliation{Synchrotron SOLEIL, L'Orme des Merisiers, Bat. A, Saint-Aubin, Gif-sur-Yvette 91192, FRANCE.}

\author{K. Medjoubi}
\affiliation{Synchrotron SOLEIL, L'Orme des Merisiers, Bat. A, Saint-Aubin, Gif-sur-Yvette 91192, FRANCE.}

\author{R. Nagaoka}
\affiliation{Synchrotron SOLEIL, L'Orme des Merisiers, Bat. A, Saint-Aubin, Gif-sur-Yvette 91192, FRANCE.}

\author{C. Herbeaux}
\affiliation{Synchrotron SOLEIL, L'Orme des Merisiers, Bat. A, Saint-Aubin, Gif-sur-Yvette 91192, FRANCE.}

\author{N. B\'echu}
\affiliation{Synchrotron SOLEIL, L'Orme des Merisiers, Bat. A, Saint-Aubin, Gif-sur-Yvette 91192, FRANCE.}

\author{P. Rommeluere}
\affiliation{Synchrotron SOLEIL, L'Orme des Merisiers, Bat. A, Saint-Aubin, Gif-sur-Yvette 91192, FRANCE.}

\author{A. Somogyi}
\affiliation{Synchrotron SOLEIL, L'Orme des Merisiers, Bat. A, Saint-Aubin, Gif-sur-Yvette 91192, FRANCE.}

%


\author{O. Chubar}
\affiliation{Brookhaven National Lab, PO Box 5000, Upton, NY 11973-5000, USA}

\author{C. Kitegi}
\affiliation{Brookhaven National Lab, PO Box 5000, Upton, NY 11973-5000, USA}


\author{M.-E. Couprie}
\affiliation{Synchrotron SOLEIL, L'Orme des Merisiers, Bat. A, Saint-Aubin, Gif-sur-Yvette 91192, FRANCE.}

\date{\today}

\begin{abstract}
Short period, high field undulators are used to produce hard X-rays on synchrotron radiation based storage ring facilities of intermediate energy and enable short wavelength Free Electron Laser. Cryogenic permanent magnet undulators take benefit from improved magnetic properties of $RE_2Fe_{14}B$ (Rare Earth based magnets) at low temperatures for achieving short period, high magnetic field and high coercivity. Using $Pr_2Fe_{14}B$ instead of $Nd_2Fe_{14}B$, which is  generally employed for undulators, avoids the limitation caused by the Spin Reorientation Transition phenomenon, and simplifies the cooling system by allowing the working temperature of the undulator to be directly at the liquid nitrogen one (77 K). We describe here the development of a full scale (2 m), 18 mm period $Pr_2Fe_{14}B$ cryogenic permanent magnet undulator (U18). The design, construction and optimization, as well as magnetic measurements and shimming at low temperature are presented. The commissioning and operation of the undulator with the electron beam and spectrum measurement using the Nanoscopmium beamline at SOLEIL are also reported. 
\end{abstract}



\maketitle


\section{\label{sec:level1}Introduction}

Accelerator based X-ray sources produce nowadays very intense radiation in a broad spectral range \cite{CouprieCRPhysique2008, CouprieJESRP2014}. Third generation synchrotron radiation light sources, with reduced emittance and large use of insertion devices provide a high brilliance with partial transverse coherence for users, enabling for example coherent imaging experiments \cite{cloetens1996phase}. Furthermore, fourth generation light source generally rely on the Free Electron Laser (FEL) process using relativistic electrons propagating in a periodic magnetic field as a gain medium. FELs provide additional longitudinal coherence and extremely short pulses, enabling to follow dynamics process of dilute species \cite{gaffney2007imaging}. 
\\
A planar undulator of period $\lambda_u$  and peak field $B_0$ emits with an observation angle $\theta$ a radiation at wavelength $\lambda_r$ and its harmonics according to: 
\begin{equation}
\lambda_r = \frac{\lambda_u}{2\gamma^2} (1+\frac{K^2}{2} + \gamma^2 \theta^2) 
\label{Eqn1}    
\end{equation}
with the deflection parameter given by $K = (eB_0\lambda_u)/(2mc) = 0.0934 \times B_0[T] \times \lambda_u[mm]$, $e$ is the electron charge, $m$ the electron mass,  $c$ the speed of light, and $\gamma$ the normalized energy. The radiation produced in an undulator is very intense and concentrated in narrow energy bands in the spectrum. 
$\lambda_r$ can be varied by a modification of the undulator magnetic field amplitude (Eq. \ref{Eqn1}). Undulators with high fields and short period enabling a higher number of periods to be used for a given length can be employed on intermediate energy storage rings \cite{CouprieJESRP2014} and  short wavelength FEL, for providing hard X-ray radiation. 
\\
Different technologies can be used for generating the undulator periodic magnetic structure \cite{elleaume2000design, couprie2012radiation}. In the case of a permanent magnet undulator, in the  so-called \textit{Halbach} configuration \cite{halbach1981physical}, two parallel arrays separated by an air-gap accommodate magnets with the magnetisation rotating from one block to the other by 90 $^\circ$. In hybrid undulators \cite{halbach1983permanent}, the vertical magnetized magnet blocks are replaced by soft iron poles which further increase the magnetic field strength in the gap of the undulator.\\
The undulator peak field can be enhanced as the magnetic gap between the upper and lower arrays of magnets is decreased. The reduction of the gap is limited by the size of the vacuum chamber  \cite{gudat1986undulator} and sets some restrictions in terms of physical aperture for the electron beam evolution. The idea to place the entire undulator in vacuum enabled the user to reach smaller gaps, and thus higher magnetic fields  \cite{yamamoto1992construction}.

Rare earth materials are used for the magnets to generate the magnetic field \cite{OSheaPRSTAB2010}. Large coercive materials have better resistance against demagnetisation that might occur due to electron beam losses or vacuum baking \cite{bizen2003baking}. $Sm_2Co_{17}$  \cite{ChavannePAC2003} magnets present a high coercivity $H_{cj}$ and a good radiation resistance against demagnetisation \cite{HaraAPAC2004}. $Nd_2Fe_{14}B$ 
\cite{kitamura2000recent}
magnets achieve a higher remanent field $B_r$, with  intermediate coercivities \cite{bizen2003baking}. Unfortunately large coercive $Nd_2Fe_{14}B$ magnets show small remanence magnetization. Therefore, the undulators can not take full advantage of the magnetic performance of $Nd_2Fe_{14}B$. In order to shift further the emitted radiation towards higher energies; i.e. to the hard X-ray region, the peak magnetic field of the in-vacuum undulators can be increased when operating at cryogenic temperature. Cooling down $Nd_2Fe_{14}B$ permanent magnets increases the remanent magnetization $M_r$ up to a certain temperature at which the process is limited by the appearance of the Spin Reorientation Transition (SRT) phenomenon \cite{GivordSSC1984, garcia2000orbital}. The easy magnetisation axis is tilted from the crystallographic c-axis [001] by an angle that increases when lowering the temperature \cite{hirosawa1986magnetization}. The coercivity is not affected by the SRT phenomenon and remains increasing at low temperature. In contrast to the $Nd_2Fe_{14}B$  case, when a $Pr_2Fe_{14}B$ permanent magnet is cooled down to cryogenic temperature, no SRT occurs  and its remanent magnetization keeps increasing at least until the liquid helium temperature of 4.2 K \cite{hiroyoshi1987high, goll1998magnetic}. Such a magnet grade is well adapted for the cryogenic undulator application since it enables direct cooling at the liquid nitrogen temperature (77 K), enabling a high level of  thermal stability.

Cryogenic Permanent Magnet Undulator (CPMU) is one of the evolutions of in-vacuum undulators. Initially proposed at SPring-8 \cite{hara2004cryogenic}, where a prototype of 0.6 m length with a period of 14 $mm$, using high remanence $Nd_2Fe_{14}B$ grade cooled down to 140 $K$, has been developed. 
Following the first demonstration at SPring-8  \cite{hara2004cryogenic, tanaka2007magnetic}, several CPMU prototypes were built. After first prototypes \cite{tanabe2007x, tanabe2010cryogenic}, a 8 $mm$ x 14.5 $mm$ periods CPMU prototype has been built at Brookhaven National Laboratory (BNL) using $Nd_2Fe_{14}B$ grade reaching a magnetic gap of 5 $mm$. Another 8 periods prototype with 16.8 $mm$ period and a gap of 5 $mm$ hybrid undulator composed of high coercive $Pr_2Fe_{14}B$ magnets (CR-47) and Vanadium Permendur poles, have been manufactured at BNL \cite{kitegi2012development} in order to investigate the resistance of the CR-47 magnets to a baking at 373 $K$. Helmholtz-Zentrum Berlin (HZB) with the collaboration of UCLA \cite{bahrdt2010cryogenic}, built two CMPU prototypes (20 x 9 $mm$ period and magnetic gap of 2.5 $mm$), using $Pr_2Fe_{14}B$ magnets cooled down to a of 20-30 $K$. At SOLEIL, three hybrid prototypes CPMU have been built and characterized, a 4 periods 18 $mm$ length with $Nd_2Fe_{14}B$ magnets (BH50 Hitachi-Neomax)\cite{BenabderrahmaneNIM2012}, also a 4 $mm$ x 18 $mm$ period with $Pr_2Fe_{14}B$ (CR53 Hitachi-Neomax) \cite{couprie2014status}, and a 4 $mm$ x 15 $mm$ period with $Pr_2Fe_{14}B$.

The first full scale cryogenic undulator had been developed at ESRF \cite{KitegiEPAC2006} with a period length of 18 $mm$ using a relatively low remanence $Nd_2Fe_{14}B$ magnet grade ($B_r$=1.16 $T$) cooled down to around 150 K, reaching a gap of 6 $mm$. ESRF recently developed two more cryogenic undulators, one with the same period length as the first using a high remanence $Nd_2Fe_{14}B$ ($B_r$=1.38 $T$) cooled down to 135 $K$ \cite{ChavanneEPAC2008}, and a second one of period 14.5 $mm$ using $Pr_2Fe_{14}B$ magnet grade cooled down to 100 $K$. Based on the ESRF development, Danfysik build for Diamond \cite{OstenfeldIPAC2010} a 17.7 $mm$ period cryogenic undulator using high remanence $Nd_2Fe_{14}B$ magnets cooled down to 150 K and a magnetic gap of 5 $mm$. $Pr_2Fe_{14}B$ based CMPUs are under construction at Diamond. 
SPring-8 in collaboration with SLS \cite{TanakaIPAC2010} developed a cryogenic undulator using high remanence $Nd_2Fe_{14}B$ cooled down around 140 K. The cryogenic undulators developed so far use $Nd_2Fe_{14}B$ permanent magnet working around 140 K. They are cooled down to the liquid nitrogen temperature at 77 K and heated to reach the working temperature in order to avoid the SRT phenomenon appearance. HZB  is currently building 2 new $Pr_2Fe_{14}B$ based CMPU (1.6 $m$ long with a 17 $mm$ period with a gap of 5 $mm$, and a 2 $m$ long with 15 $mm$ period and a gap of 2 $mm$). A 2 $m$ long cryogenic undulator with 140 $mm$x 13.5 $mm$ period is under construction for the High Energy Photon Source Test Facility (HEPS-TF) in Korea. $Pr_2Fe_{14}B$ based CMPUs are under construction at National Synchrotron Radiation Research Center \cite{huang2016challenge}.

In this paper, the development of the first full scale (2 m long) $Pr_2Fe_{14}B$ cryogenic undulator which has been installed on a storage ring is presented, i.e. SOLEIL in France. It has 107 periods of length 18 mm. The magnetic analysis and measurement results indicate that the quality of the magnet grade satisfies the undulator requirements and demonstrate a further increase of magnetic field (by a few percents). We show that the direct operation at 77 K enables a good thermal stability. We also present the design, construction steps (with optimization, magnetic measurements and shimming at low temperature). We then report on the commissioning with the electron beam and the current successful operation both from an electron or photon point of view (with measurements of undulator radiation). Indeed, we show that some results of photon beam alignment using the Nanoscopium long section beamline, or precise adjustment of the undulator taper with the photon beam itself.

\section{\label{sec:level2}Design of the cryogenic undulator}

\subsection{\label{sec:level21}Magnetic design}

\begin{figure}[!ht]
  	\centering
    \includegraphics[scale=0.8]{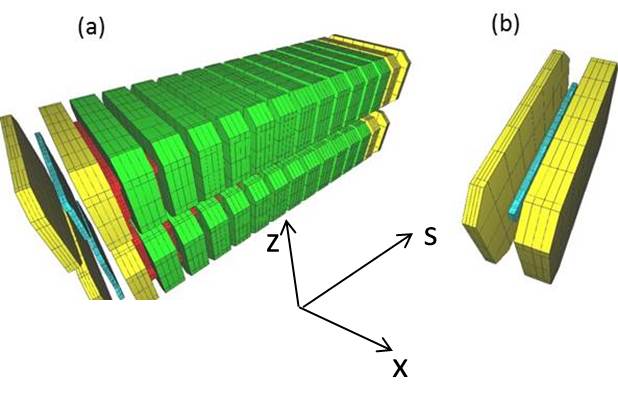}
    \caption{Magnetic design using RADIA \cite{ChubarJSR1998}.  (a) Seven period undulator : permanent magnet (green) of the main undulator part ($50 \times 30 \times 6.5$ $mm^3$), Poles of the main undulator part (red) ($33 \times 22 \times 2.5$ $mm^3$),   permanent magnet (yellow) of the extremity part of the undulator($80 \%$ and $35 \%$ of the main magnet width)  \cite{ChavannePAC1999}. Pole of the extremity part  (blue) of the undulator ($50 \%$ of the main polewidth). (b) Extremity part of the magnetic design. }
    \label{Fig1}
\end{figure}

The SOLEIL $Pr_2Fe_{14}B$ based U18 cryogenic undulator has been modeled using RADIA \cite{ChubarJSR1998} software, as illustrated in Fig. \ref{Fig1}. The characteristics are indicated in Table \ref{Table2}, considering the longitudinal coordinate s, the transverse horizontal and vertical are x and z respectively. The model contains two parts; the central one (fig. \ref{Fig1}(a)) constituted by full size permanent magnets and poles producing the main magnetic field of the undulator, and the extremities (fig. \ref{Fig1}(b)) located at each end of the undulator constituted by two magnets and one pole. The extremity parts are optimized to minimize the on-axis field integral variations versus the undulator gap \cite{ChavannePAC1999}. 

\begin{table}[h]
	\scriptsize
	\caption{SOLEIL Cryogenic undulator main characteristics.}
	\centering
		\begin{tabular}[c]{lcc}
		\hline
		\textbf{Item}&\textbf{Unit}&\textbf{Value}\\
		\hline
		\textbf{Technology}&&Hybrid\\
		\textbf{Magnet Material CR53 (Hitachi)}&&$Pr_2Fe_{14}B$\\
		\textbf{Remanence $B_r$}&T&1.35 at 293 K\\
		&&1.57 at 77 K\\
		\textbf{Coercivity $H_{cj}$}&T&1.63 at 293 K\\
		&&7.6 at 77 K\\
		\textbf{Magnet size (x, z, s)}&$mm^3$&$50 \times 30 \times 6.5$\\
		\textbf{Pole material}&&Vanadium Permandur\\
		\textbf{Pole size(x, z, s)}&$mm^3$&$33 \times 22 \times 2.5$\\
		\textbf{Period}&mm&18\\
		\textbf{Minimum magnetic gap}&mm&5.5\\
		\textbf{Maximum magnetic gap}&mm&30\\
		\textbf{Magnetic peak field at minimum gap}&T&1.152\\
		\textbf{Deflection parameter}& &1.936\\
		\textbf{Number of periods}&&107\\
		\hline \hline
		\end{tabular}
	\label{Table2}
\end{table}

\begin{figure}[!ht]
  	\centering
    \includegraphics[scale=0.4]{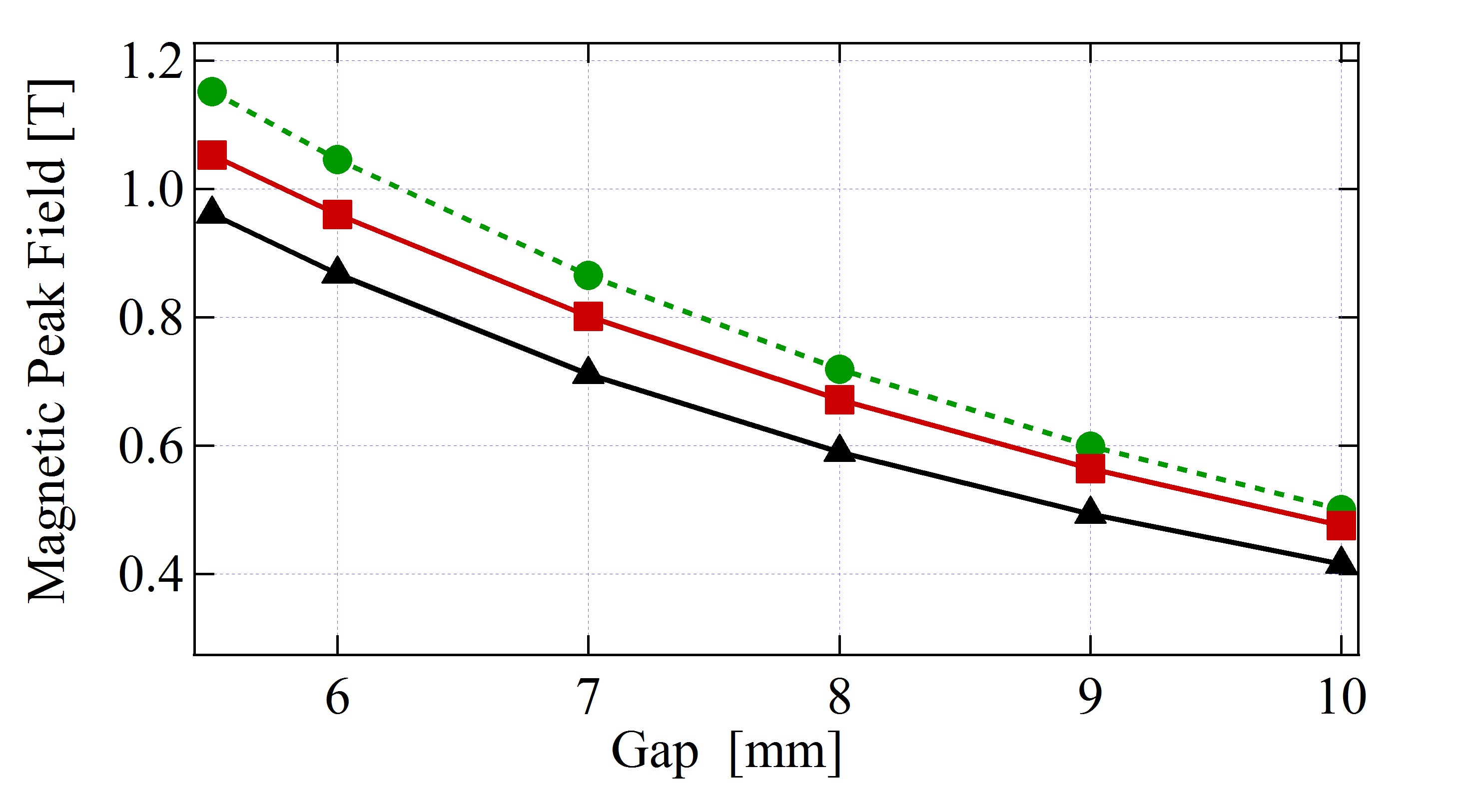}
    \caption{On-axis magnetic peak field calculated with RADIA. {(\color{green} $\bullet$}) U18 $Pr_2Fe_{14}B$ cryogenic undulator at 77 K, $B_r$ = 1.57 T at 77 K, magnet dimensions 50 $mm \times$ 30 $mm \times$ 6.5 $mm$, pole dimensions 33 $mm \times$ 22 $mm \times$ 2.5 $mm$. ({\color{red}$\blacksquare$}) U20 $Nd_2Fe_{14}B$ in-vacuum undulator  at 293 $K$, $B_r$ = 1.25 T at 293 K, magnet dimensions 50 $mm \times$ 30 $mm \times$ 7.5 $mm$, pole dimensions 33 $mm \times$ 22 $mm \times$ 2.5 $mm$. ($\blacktriangle$) U20 $Sm_2Co_{17}$  in-vacuum undulator  at 293 $K$, $B_r$ = 1.05 $T$ at 293 $K$, magnet dimensions 50 $mm$ $\times$ 30 $\times$ 7.5 $mm$, pole dimensions 33 $mm$ $\times$ 22 $mm$ $\times$ 2.5 $mm$.}
    \label{Fig2}
\end{figure}

For a given deflection parameter and total undulator length, the increase of the undulator magnetic field allows  for the reduction of the period, so the number of periods could be enhanced, resulting in higher flux and brilliance. Figure \ref{Fig2} presents the on-axis magnetic peak field versus gap calculated for cryogenic undulators U18 ($Pr_2Fe_{14}B$), U20 ($Nd_2Fe_{14}B$, and $Sm_2Co_{17}$). The magnetic field of U18 cryogenic undulator cooled down to 77 K is $\sim 10\%$ higher than U20 $Nd_2Fe_{14}B$ based undulator and $\sim 20\%$ higher than U20 $Sm_2Co_{17}$ at room temperature.

\subsection{\label{sec:level21}Expected spectral properties}

In the case of SOLEIL, the 18 $mm$ period is chosen to keep the deflection parameter K close to the value used for U20 in-vacuum undulators one. The  number of periods to be assembled can be increased by $11\%$  with the same undulator length, which enhances the brightness for the same photon energy spectrum range.

\begin{figure}[!ht]
\centering
\includegraphics[scale=0.4]{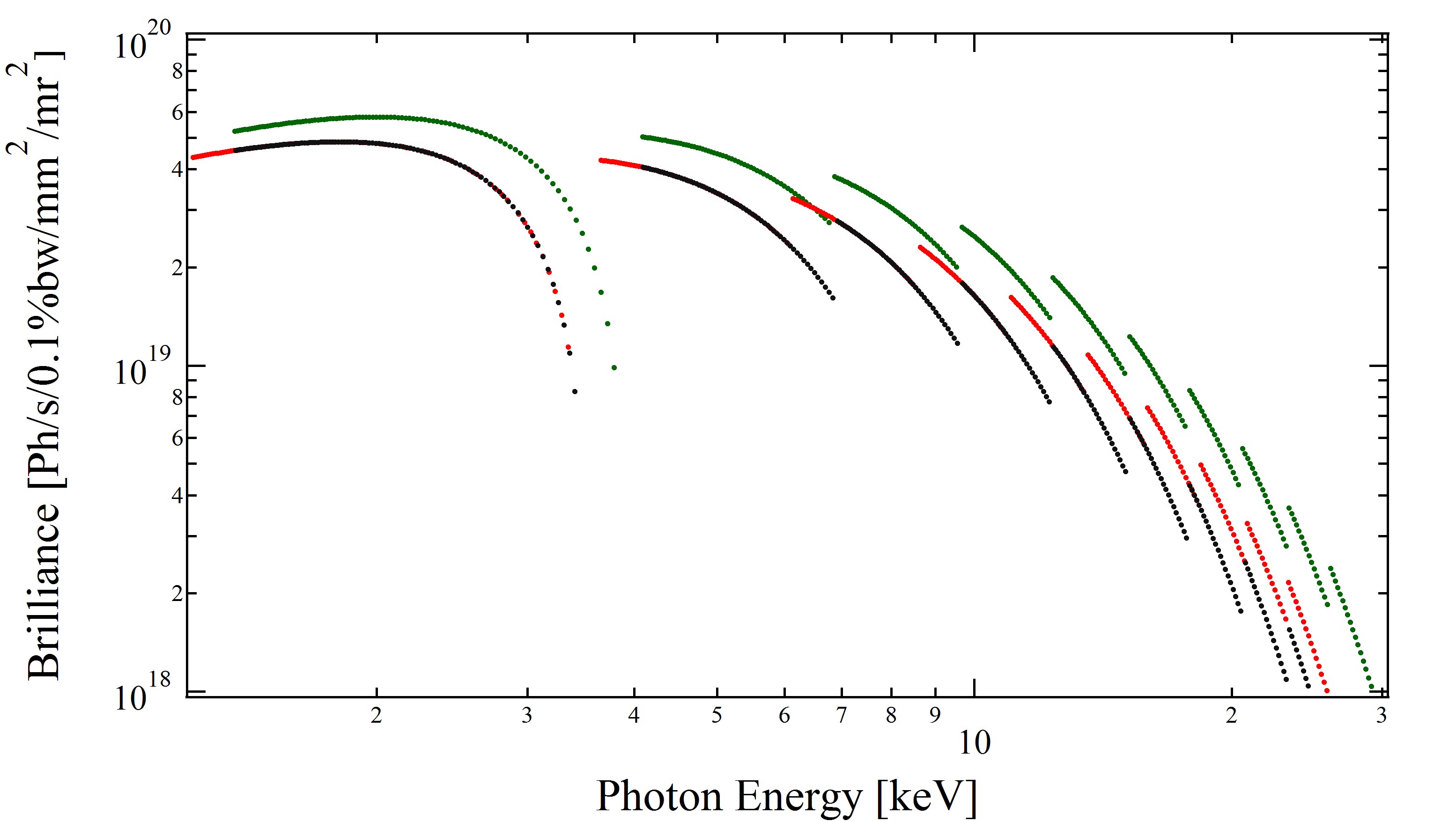}
\caption{Brilliance in logarithmic scale with gap= 5.5 $mm$ calculated with SRW  \cite{ChubarEPAC1998} for three different undulators: (Black) U20 $Sm_2Co_{17}$ with peak field of 0.971 $T$, (Red) $Nd_2Fe_{14}B$ with peak field of 1.053 $T$,  and (Green) $Pr_2Fe_{14}B$ with peak field of 1.152 $T$ at cryogenic temperature. Electron beam characteristics of Table \ref{Table12}, with $\beta_x$= 5.577	$m$, $\beta_z$= 8.034 $m$, $\alpha_x$= 0, $\alpha_z$= 0.001 $rad$), through a window aperture of 0.1 $mm$ $\times$ 0.1 $mm$ placed at a distance of 11.7 $m$.}
\label{Fig:Brilliance}
\end{figure}

\begin{table}[h]
	\small
	\caption{Characteristics of the electron beam in SOLEIL long section beamline}
	\centering
		\begin{tabular}[c]{ccc}
		\hline
		\textbf{Parameters}&\textbf{Value}&\textbf{Unit}\\
		\hline
Energy	&2.75&	GeV\\
Current	&0.5&	A\\
Emittance H	&3.9&	nm.rad\\
Emittance V	&0.039&	nm.rad\\
RMS energy spread	&0.1016&	$\%$\\
		\hline \hline
		\end{tabular}
	\label{Table12}
\end{table}

\begin{figure}[!ht]
  	\centering
    \includegraphics[scale=0.5]{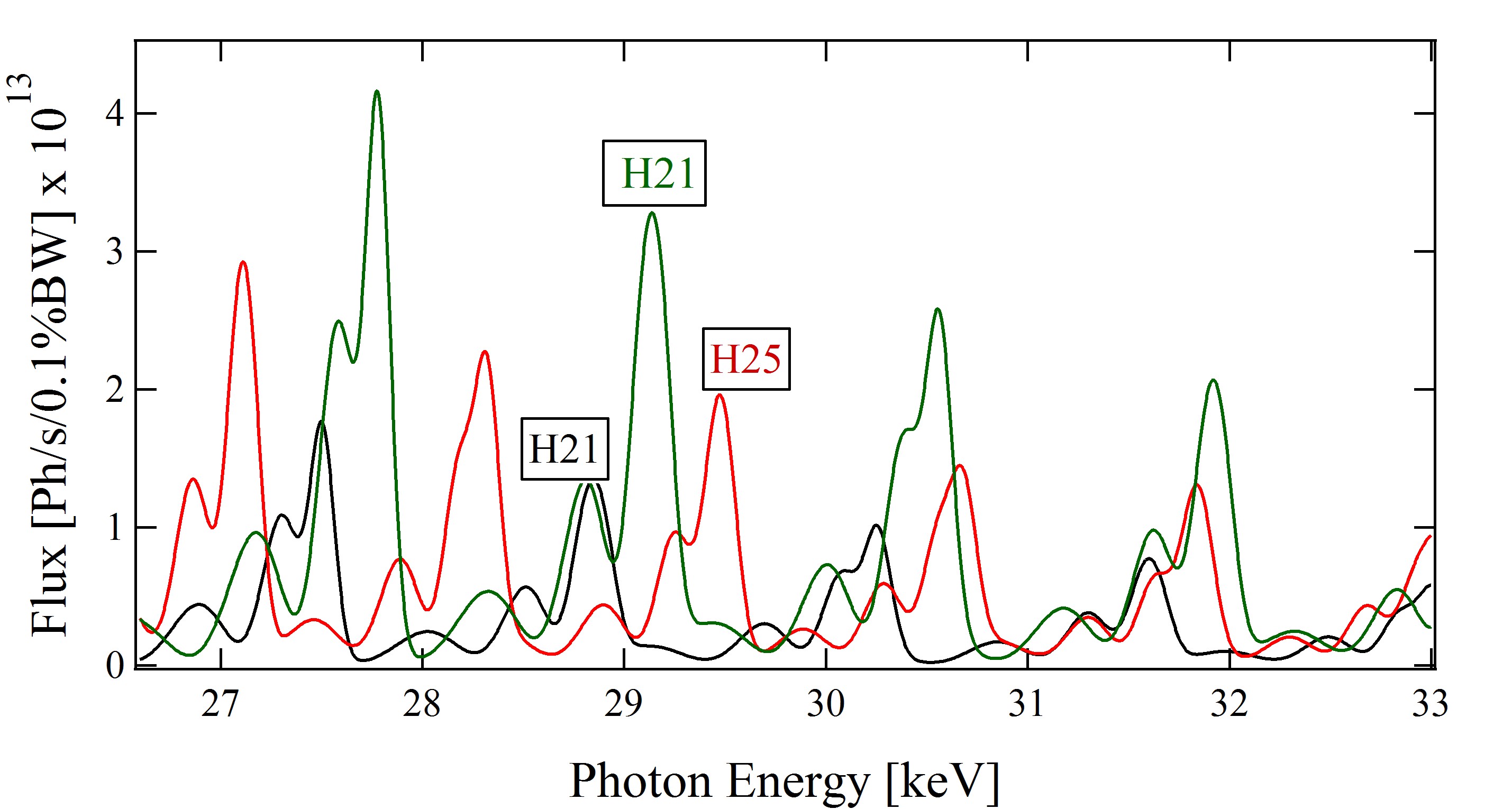}
    \caption{Spectrum calculated with SRW with observation window at 11.7 $m$ through a slit 0.1 $mm$ $\times$ 0.1 $mm$ at minimum gap of 5.5 $mm$. Beam characteristics shown in Table \ref{Table12}, with $\beta_x$= 5.577	$m$, $\beta_z$= 8.034 $m$, $\alpha_x$= 0, $\alpha_z$= 0.001 $rad$. (Green) U18 $Pr_2Fe_{14}B$ cryogenic undulator at 77 $K$, $B_{z0}$ = 1.152 T, K = 1.937 . (Red) U20 $Nd_2Fe_{14}B$ in-vacuum undulator at 293 K, $B_{z0}$ = 1.053 $T$, K = 1.967. (Black) U20 $Sm_2Co_{17}$ in-vacuum undulator at 293 K, $B_{z0}$ = 0.971 $T$, K = 1.812. H with the index represents the harmonic number at photon energy $\sim$ 29 $keV$. }
    \label{Fig3}
\end{figure}

The spectral performance of the $Pr_2Fe_{14}B$ cryogenic undulator U18 is compared to the in-vacuum undulator U20 $Nd_2Fe_{14}B$ and U20 $Sm_2Co_{17}$ used in SOLEIL storage ring on a long straight section (see Table II for the parameters). The period is slightly adjusted so that typically the same spectral range is covered and the total length is kept constant. 
Figure \ref{Fig:Brilliance} and figure \ref{Fig3} respectively show the brilliance and spectra calculated with SRW software \cite{ChubarEPAC1998} for the three given undulators mentioned before. Larger brilliance and flux is achieved with U18 than the two U20 undulators. At a photon energy of 30 $keV$, U18 flux is $\sim$ 2 times higher than the one of the U20 $Nd_2Fe_{14}B$ undulator and 2.5 times higher than U20 $Sm_2Co_{17}$ undulator.

In the case of cryogenic undulators, these energies are obtained at lower harmonic order than in the case of room temperature in-vacuum undulators. In consequence, there is less intensity reduction due to the undulator imperfections, since lower order harmonics are less sensitive to phase error $ \sigma_{\phi}$ \cite{walker1993interference, walker2013phase}. Indeed, phase error is due to the magnetic field errors along the undulator axis, such as variations in the peak field or the period length from one period to another. It causes change in the length of the electron trajectory and yields a phase lag $\Phi$, between the electron and the photon expressed as : $\Phi=\frac{2 \pi}{\lambda_R}\big(\frac{l(s)}{\beta}-s\big)$
where $l(s)$ is the path traveled by the electron up to the point s. The phase lag causes the interference of electrons and photons to occur on different wavelengths and add destructive interference, which results in line broadening and intensity reduction of the emitted lines. The effect is larger on higher harmonics, due to the fact that they are more sensitive to the phase slippage. Indeed, the intensity reduction can be expressed as : $I_n=I_{n0} \exp{(-n^2 \sigma_{\phi}^2})$ with n the harmonic order, and $I_{n0}$ the intensity without phase error. For example, 29 $keV$ is reached on harmonic 21 with the cryogenic U18 $Pr_2Fe_{14}B$ and harmonic 25 with U20 $Nd_2Fe_{14}B$ as shown in Fig. \ref{Fig3}.
For a $2.5^{\circ}$ phase error, the intensity reduction of  the 21 harmonic is calculated to be of factor $\sim 0.56$, as for the 25 harmonic the reduction is of the factor $\sim 0.70$. However the effect of phase errors can become less severe than predicted by the above equation when electron beam emittance and energy spread are taken into account. So with low phase error, higher harmonics with a fair intensity can be achieved.


\subsection{\label{sec:level22}Mechanical design}
\begin{figure}[!ht]
  	\centering
    \includegraphics[scale=1]{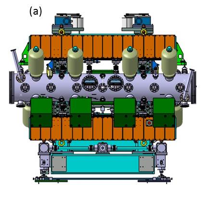}
    \caption{Mechanical design of the cryogenic undulator using 3D CATIA (http://www.3ds.com/fr/).}
    \label{Fig4}
\end{figure}

Figure \ref{Fig4} presents the mechanical design of the cryogenic undulator U18 inspired from the in-vacuum undulator \cite{ChavannePAC2001}. The carriage is constituted of a metallic base where the frame is welded. Two out-vacuum (external) girders are fixed on the frame and can move vertically thanks to two series of sliders. The magnetic systems (permanent magnets, poles and their mechanical supports) are fixed on two in-vacuum girders connected to the external ones by 24 rods. The in-vacuum girders are separated by a gap where the electron beam crosses the undulator. The gap variation (from minimum value of 5.5 $mm$ to maximum value of 30 $mm$) is enabled by two steps motors Berger Lahr VRDM3910; a third step motor is used to move vertically the undulator over a 10 $mm$ range in order to align in the vertical direction the magnetic axis with the electron beam axis. The in-vacuum girders with the magnetic system are installed in a vacuum chamber equipped with Ion pumps, Titanium sublimation pump and instrumentation to ensure an Ultra High Vacuum in the vacuum chamber during the operation with electron beam. Cu-OFHC tapers are fixed on the vacuum chamber and on the in-vacuum girders to guaranty a smooth variation of the impedance seen by the electron beam when it crosses the undulator.

A copper absorber is installed at the downstream of the undulator inside the vacuum chamber to collect the undesired photon beam coming from the upstream bending magnet. The absorber is cooled down with water at room temperature. A 100 $\mu m$ Cu-Ni foil is placed on the magnetic system and stretched at the extremities of the undulator by a spring tensioner system. It conducts the image current generated by the electron beam when it crosses the undulator at a very close position from the magnetic system. Permanent magnets and different parts of the undulator inside the vacuum chamber are equipped with several temperature sensors (thermocouples and platinum sensors PT100) in order to measure the temperature during the cooling down and storage ring operation.

The main difference between the cryogenic undulator and a room temperature in-vacuum undulator is the cooling system. In the cryogenic undulator using $Nd_2Fe_{14}B$, the in-vacuum girders are connected to the cooling pipes (cooled down to the liquid nitrogen temperature of 77 $K$) through longitudinally distributed spacers acting as thermal resistors to bring the magnetic system temperature to around 150 $K$ \cite{KitegiEPAC2006, ChavanneEPAC2008, OstenfeldIPAC2010, TanakaIPAC2010}. In contrast with $Nd_2Fe_{14}B$ based cryogenic undulators, $Pr_2Fe_{14}B$  based cryogenic undulators  are cooled down directly to 77 $K$, in which the liquid nitrogen crosses the in-vacuum girders through a 12 $mm$ diameter hole guaranteeing a better temperature distribution and thus a smaller thermal gradient along the magnetic system.

\begin{table}[!ht]
	\small
	\caption{Total heat load dissipated by the liquid nitrogen close loop}
	\centering
		\begin{tabular}[c]{cc}
		\hline
		\textbf{Heat load source}&\textbf{Power (W)}\\
		\hline
Vacuum Chamber	&70\\
Rods	&104\\
Electron beam	&50\\
Total power dissipated	&224\\
		\hline \hline
		\end{tabular}
	\label{Table19}
\end{table}

 Table \ref{Table19} presents the total calculated thermal heat load dissipated by the liquid nitrogen closed loop in the case of the $U18$ build at SOLEIL. The heat load comes from heat deposited by the beam (wakefield and synchrotron radiation), the vacuum chamber, and the rods. The heat load coming from the electron beam represents more than 25$\%$ of the total heat load, due to the synchrotron radiation coming from the upstream dipole and the wake field power deposited by the passage of the electron beam on the liners and tapers. Half of the dissipated heat load is coming from the rods, which connect the out-vacuum and in-vacuum girders. The diameters and contact surfaces of the rods have been reduced without any effect on their rigidity to reduce the heat load. Since the vacuum chamber is at room temperature, whereas the in-vacuum girders  are at liquid nitrogen temperature, the inside part of the vacuum chamber has been polished to improve the emissivity and to reduce the heat load by a factor of 2.

The cooling down of the undulator to liquid nitrogen temperature causes contraction on the in-vacuum girders and the rods, calculated with equation $\Delta L= \alpha . L. \Delta T$, where $\Delta L$ is the contraction, $\alpha$ is the thermal dilatation coefficient and $\Delta T$ is the temperature variation. The contraction of the in-vacuum girder by 8 mm is then taken into account in the magnet assembly at room temperature by keeping 40 $\mu m$ between magnet modules. 
The contraction of the rods by 1 mm causes an opening of the gap by 1 mm, which should be adjusted before conducting magnetic measurements.

\section{\label{sec:level5}Assembly, magnetic optimization and measurement of the Cryogenic undulator}

\subsection{\label{sec:sublevel1}Assembly and optimization at room temperature}

The assembly and the magnetic corrections of the cryogenic undulator are performed at room temperature in the same conditions as a standard in-vacuum undulator. A standard magnetic bench allowing Hall probe and flip coil measurements has been used for the assembly and corrections at room temperature. An optimization software called ��ID-Builder�� \cite{ChubarAIPCP2007} developed at SOLEIL has been used at all steps of the undulator construction: magnets sorting, period assembly, shimming (vertical displacement of magnets and poles to correct the field integrals and the phase error), and magic fingers (small magnets installed at the extremities of the undulator to correct the field integral). The electron beam should emit the most intense radiation when it crosses the undulator without disturbing the beam dynamics in the storage ring. The figures of merit during the assembly and corrections are the field integrals, the trajectory straightness and the phase error. They have to be minimized to reduce the impact of the magnetic errors on the undulator performances in terms of photon spectrum and beam dynamics.

\begin{figure}[!ht]
  	\centering
    \includegraphics[scale=.4]{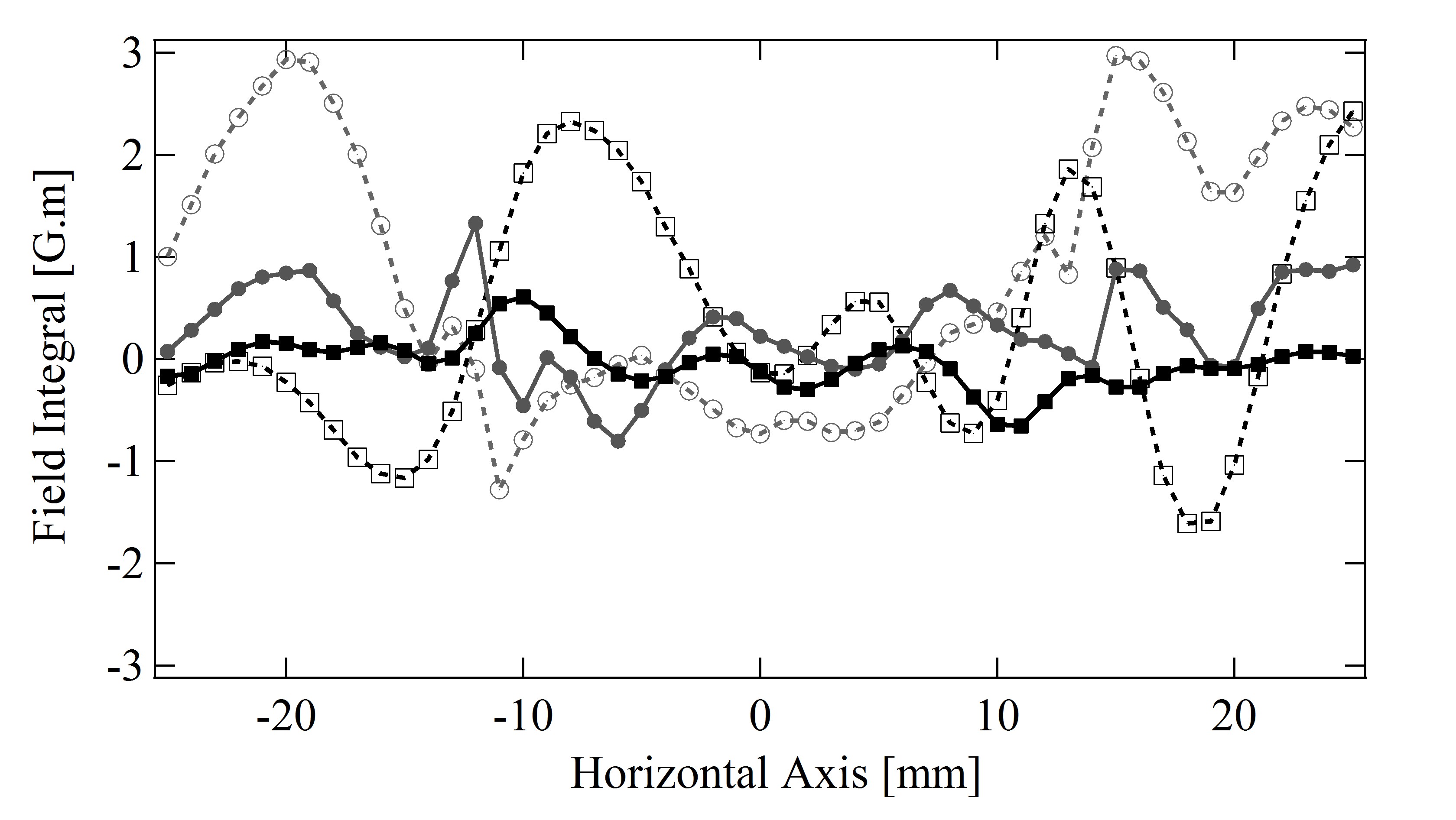}
    \caption{Field integrals versus horizontal position at minimum gap of 5.5 mm at room temperature. ($\blacksquare$) Vertical field integral (Iz) after magic finger corrections. ($\bullet$) Horizontal field integral (Ix) after magic finger corrections. ($\square$) Vertical field integral (Iz) before magic finger corrections ($\circ$) Horizontal field integral (Ix) before magic finger corrections. Field integrals precision is $\sim$ 0.05 $G.m$, and local measurements (Hall probe) is 0.5 $G$.}
    \label{Fig5}
\end{figure}

Figure \ref{Fig5} presents the vertical field integral $I_z$ and the horizontal field integral $I_x$ of the cryogenic undulator versus transverse position at minimum gap of 5.5 mm at the end of the assembly at room temperature and after the complete iterative correction process. After the corrections, the undulator field integrals present a smoother variation versus horizontal position and the on-axis integral is less than 0.4 $G.m$ and the higher off-axis field integral are reduced from 3 $G.m$ to less than 1 $G.m$.

\begin{figure}[!ht]
  	\centering
    \includegraphics[scale=.4]{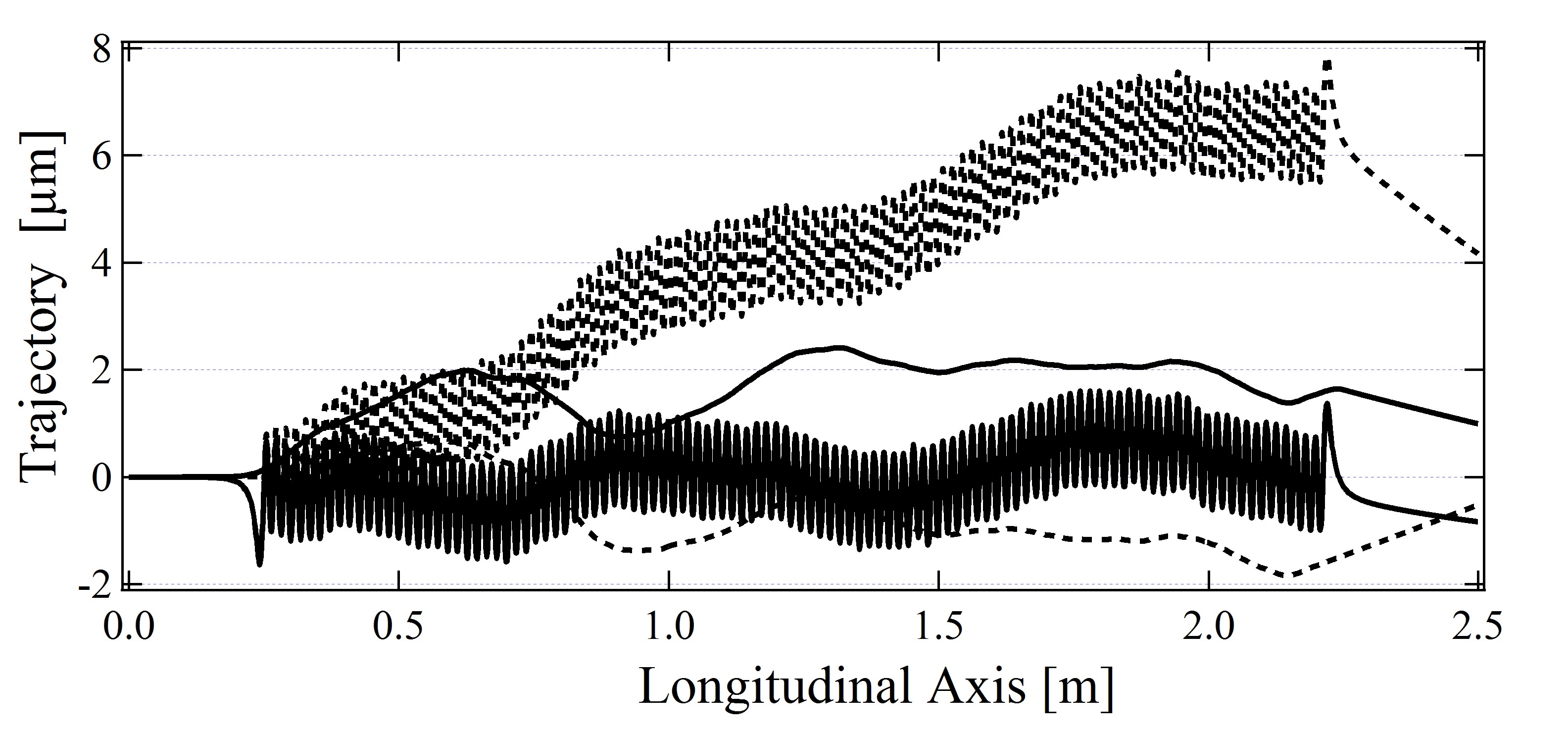}
    \caption{Electron beam trajectory in the cryogenic undulator at room temperature calculated with B2E code from Hall probe magnetic measurement at SOLEIL (E= 2.75 $GeV$). (Solid line) Horizontal and vertical trajectories after the assembly and corrections. (Dash line) Horizontal and vertical trajectories after the assembly and before shimming.}
    \label{Fig6}
\end{figure}

Figure \ref{Fig6} presents the electron beam trajectory in the cryogenic undulator calculated from the magnetic field measurements versus the longitudinal position at the end of the assembly at room temperature and after the corrections. The trajectory position at the exit of the undulator after the magnetic assembly is 8 $\mu m$, it is kept below 2 $\mu m$ after magnetic corrections.

The RMS phase error (calculated by B2E  from the measured magnetic field \cite{elleaume1991b2e}) at minimum gap after the magnetic assembly of the undulator is found to be $14.5^o$, and has been corrected to $2.8^o$ after shimming the magnet modules. The vertical position of the magnet modules is modified by using shims (copper pieces) to correct locally the magnetic field value.

\subsection{\label{sec:sublevel2}Cryogenic magnetic measurement bench}

Magnetic measurement benches have been developed over the past years to increase precision and enable to conduct measurements for insertion devices at low temperature. A cryogenic undulator requires a specific bench with certain characteristics. ESRF \cite{ChavanneEPAC2008} had developed a magnetic bench to perform magnetic field measurements for cryogenic undulators. This bench is mounted inside a vacuum chamber, where the Hall probe (with a ball bearing) is fixed on a mobile carriage guided along a rail. The longitudinal position of the Hall probe is measured with a laser tracker. The stretched wire motion is performed by two translation motors located outside the vacuum chamber and transmitted inside by rods through bellows. 
SPring-8 developed a bench called ��SAFALI�� \cite{TanakaPRSTAB2009} for Hall probe magnetic measurements and is mounted inside the vacuum chamber of the undulator. The Hall probe is displaced longitudinally by a vacuum compatible stepper motor installed inside the vacuum chamber. The mechanical default position of the Hall probe on the rail is measured with lasers and corrected with small displacements on the rail. Helmholtz-Zentrum Berlin (HZB) built a 2 $m$ long in-vacuum Hall-probe measuring bench \cite{kuhn2013hall} for the characterization of several in-vacuum cryogenic undulators under development. The bench employs a system of laser interferometers  and position sensitive detectors, which are used in a feed-back loop for the Hall probe position and orientation.

\begin{figure}[!ht]
  	\centering
    \includegraphics[scale=.81]{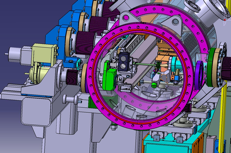}
    \caption{SOLEIL dedicated magnetic measurement bench inside the vacuum chamber of the cryogenic undulator.}
    \label{Fig7}
\end{figure}

Figure \ref{Fig7} presents the cryogenic magnetic measurement bench developed at SOLEIL. It is used to measure the field created by the undulator at both room and cryogenic temperature. The local magnetic field and field integrals are measured by a Hall probe (model Bell GH701) and a stretched wire respectively. The cryogenic magnetic measurement bench is installed inside the undulator vacuum chamber and is removed after the measurements. The Hall probe is fixed on a trolley which moves on a rail by a stepper motor, mounted outside the vacuum chamber and whose movement is transmitted inside the vacuum chamber via a magnetic coupling system. The rail is mechanically independent of the vacuum chamber and is fixed from the outside by seven rods. The deformations of the rail are measured with an optical system and the longitudinal position of the Hall probe is measured with a thermalized optical rule (Heidenhain LIDA 405). The deformations of the rail are measured with an optical system constituted by an angular reflector for the Hall probe angular deformation measurements, and a linear reflector (Renishaw) for the Hall probe horizontal and vertical deformation measurements.  
 

The cryogenic magnetic bench is aligned inside the vacuum chamber. The angular deformations of the bench are corrected by using shims with different thicknesses installed at the deformation location, however the vertical deformation of the bench is not corrected mechanically by shims, and it is corrected directly on the magnetic measurement results.


\begin{figure}[!ht]
  	\centering
    \includegraphics[scale=.5]{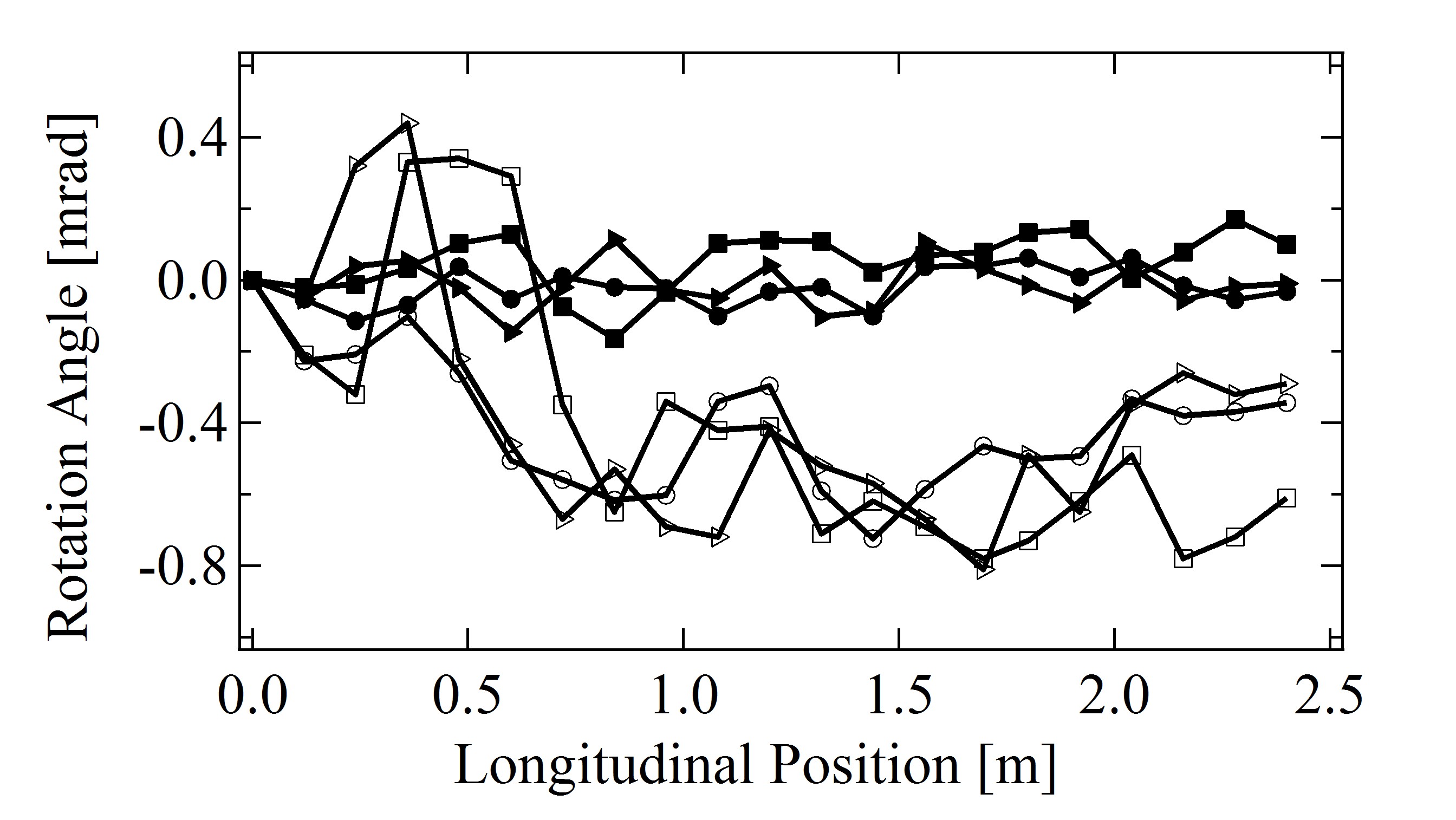}
    \caption{Hall probe rotation angle versus longitudinal position. ($\bullet$) $\theta_s$ after corrections, ($\blacksquare$) $\theta_z$ after corrections, ($\blacktriangleright$) $\theta_x$ after corrections, ($\circ$) $\theta_s$ before corrections,  ($\square$) $\theta_z$ before corrections ($\triangleright$) $\theta_x$ before corrections.}
    \label{Fig8}
\end{figure}

Figure \ref{Fig8} presents the rotation angle of the Hall probe at different longitudinal positions before and after correction. The angles are measured with a laser interferometer, they are reduced from  $\sim$0.8 $mrad$ to $\sim$0.1 $mrad$ using small mechanical shims located along the longitudinal position.

\begin{figure}[!ht]
  	\centering
    \includegraphics[scale=.5]{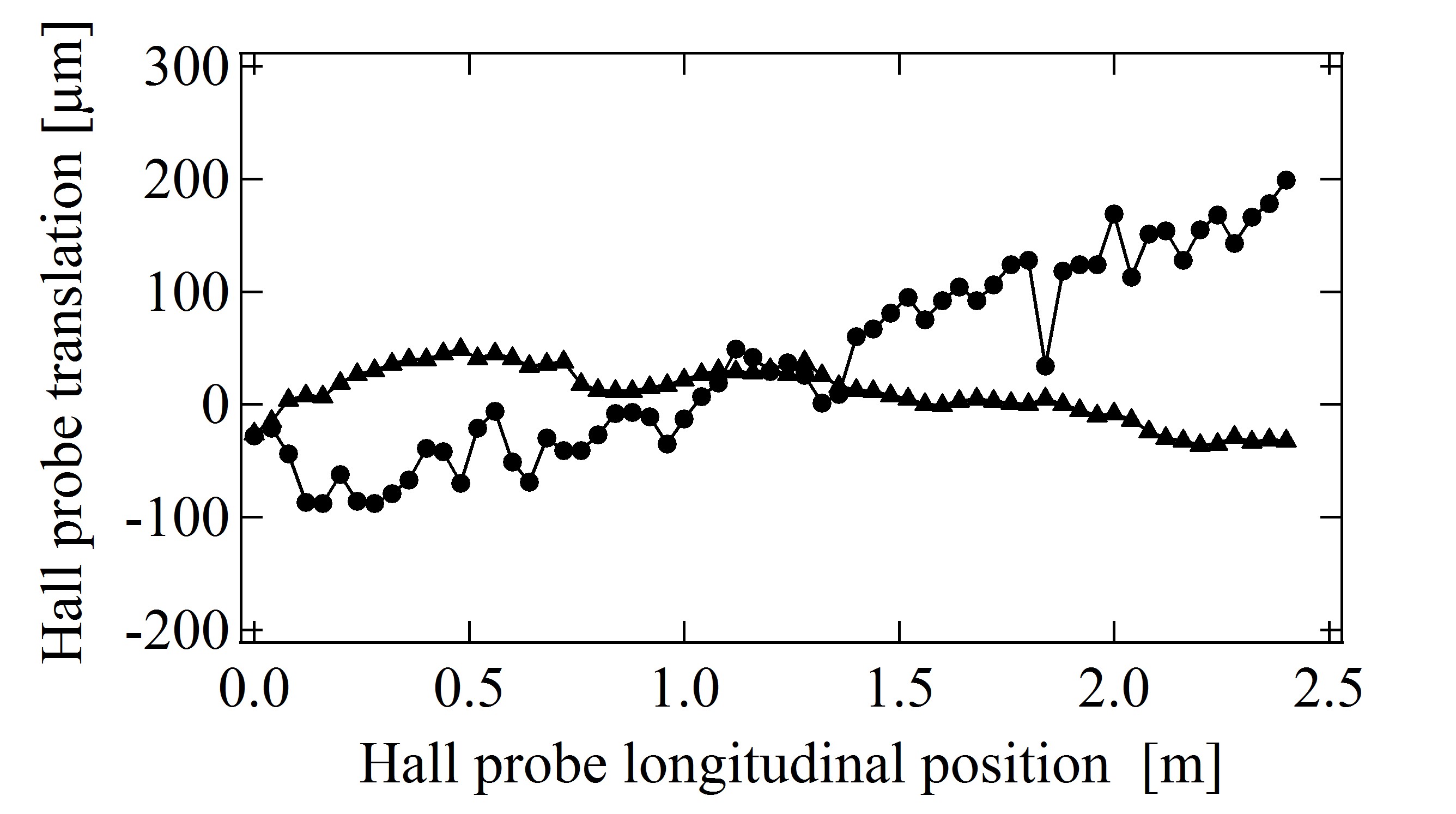}
    \caption{Hall probe translation versus longitudinal position. ($\bullet$) Translation in the vertical z axis direction,($\blacktriangle$) Translation in the horizontal x axis direction.}
    \label{Fig9}
\end{figure}

Figure \ref{Fig9} presents the measured horizontal and vertical Hall probe angles at different longitudinal positions of the cryogenic magnetic bench. The horizontal Hall probe position variation is 81 $\mu$m along the bench. This variation is negligible for the magnetic field measurement because of the very low field gradient (calculated with RADIA) over few millimeters in this direction ($10^{-8}$ $T/m$). However the vertical variation of 287 $\mu$m along the bench has a non-negligible effect on the magnetic field measurement as the field gradient (calculated with RADIA) is high in this direction (45 $T/m$). The measured magnetic field is then corrected using the equation \ref{eqn5}.

\begin{equation}
B_c(s) = B_0(s) \cosh\left(\frac{2\pi f d_z}{\lambda_u}\right)
\label{eqn5}    
\end{equation}
where $B_c$ is the corrected magnetic field, $B_0$ the measured magnetic field, $d_z$ the vertical default position, 
$s$ the longitudinal position and $f$ is a factor which depends on the undulator characteristics, a RADIA calculation identifies a value of $f$ = 1.12.

The Hall probe is calibrated with temperature variation. The resistance of the Hall probe (R) is measured at different temperatures (T) and a linear variation with temperature is measured, leading to $R(T) = a + bT$ with 
$a = -56.36 \pm 0.379$ $\Omega$, 
$b= 0.192 \pm 0.002$ $\Omega / K$. The Hall probe temperature varies by 4 $K$ during the magnetic measurements. This variation and the thermal sensitivity coefficient of the Hall probe -0.0043 $\%$/K are taken into account to correct the measured magnetic field values.

\subsection{\label{sec:sublevel3}Cooling at cryogenic temperature}

\begin{figure}[!ht]
  	\centering
    \includegraphics[scale=.5]{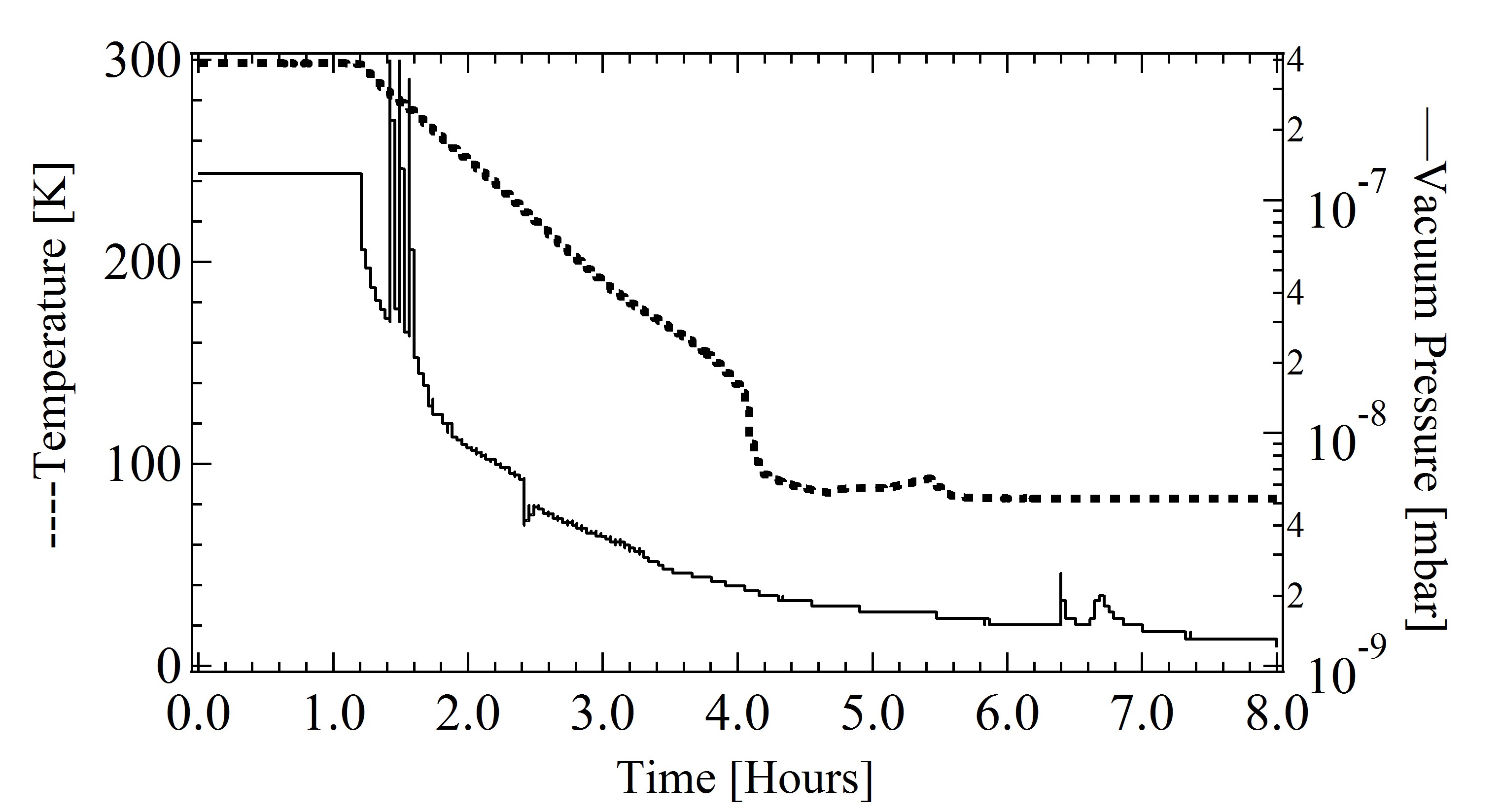}
    \caption{Cooling down of the cryogenic undulator. (normal line) Undulator vacuum pressure, (dashed line) Permanent magnets temperature.}
    \label{Fig10}
\end{figure}

Figure \ref{Fig10} presents the magnet temperature and vacuum pressure variation during the cooling down of the cryogenic undulator using a cryocooler system (Cryotherm Bruker). The magnets reach liquid nitrogen temperature of 77 $K$ after 6 hours. Despite the absence of backing of the undulator to prevent the permanent magnet from demagnetization, the undulator vacuum pressure drops quickly, due to the cold mass which acts as a cryo-pump.

\subsection{\label{sec:sublevel3}Warming up}
\begin{figure}[!ht]
\centering
\includegraphics[scale=0.4]{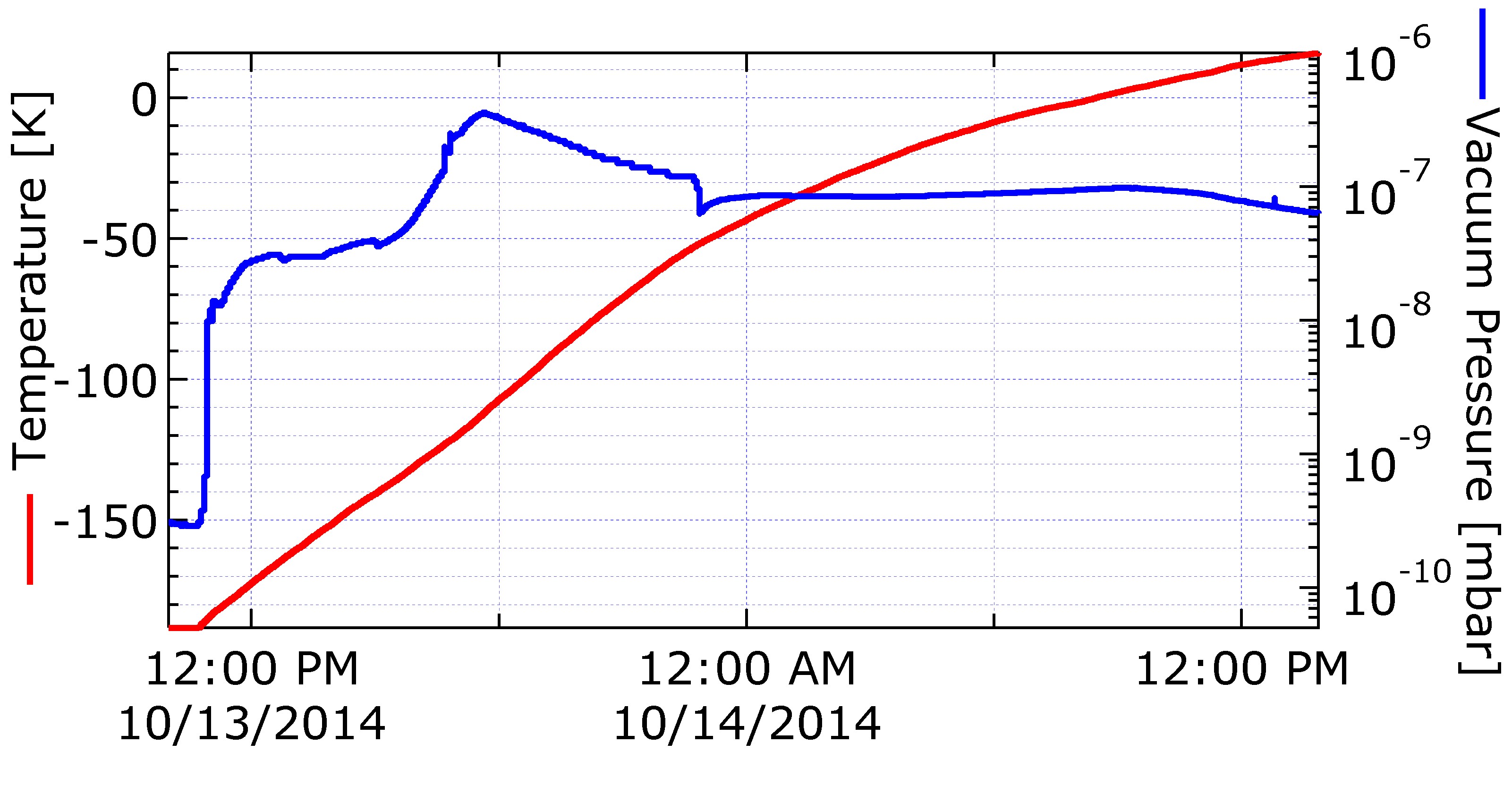}
\caption{Warming up the undulator to room temperature. (Blue) Undulator vacuum pressure, (Red) Permanent magnets temperature.}
\label{Fig:WarmingUp}
\end{figure}

The warming of the undulator from cryogenic to room temperature is relatively long, about 72 hours, but it could be accelerated (if needed) by injecting nitrogen gas in the cooling circuit. The warming time is then reduced to 36 hours. Fig. \ref{Fig:WarmingUp} shows the warming of the undulator by injecting nitrogen gas at 60 $^o$C under 2.5 $bar$ in the undulator to reduce again this long delay to approximately 24 hours.

\subsection{\label{sec:sublevel4}Measurements and magnetic measurements at cryogenic temperature}

\begin{figure}[!ht]
  	\centering
    \includegraphics[scale=.45]{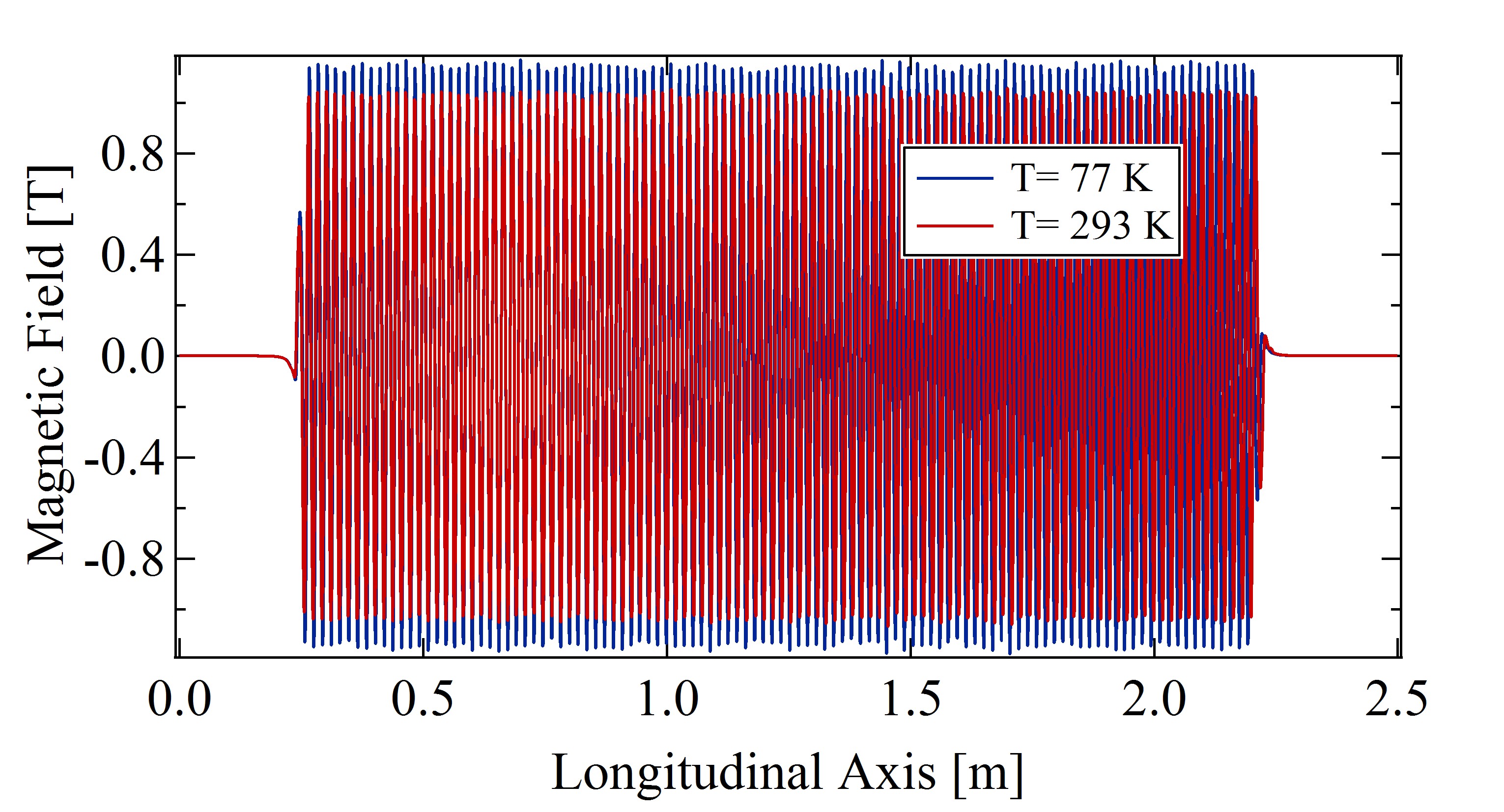}
    \caption{On axis magnetic field measurements at minimum gap of 5.5 $mm$. (Red curve) at room temperature 293 $K$, (Blue curve) at cryogenic temperature 77 $K$, with a precision of 0.5 $G$.}
    \label{Fig11}
\end{figure}

Figure \ref{Fig11} presents the Hall probe magnetic measurement results at room temperature and cryogenic temperature of 77 $K$. The effect of cooling down the permanent magnets results in an increase of the magnetic field by 11$\%$ as predicted (from 1.041 $T$ to 1.157 $T$).

\begin{figure}[!ht]
  	\centering
    \includegraphics[scale=.35]{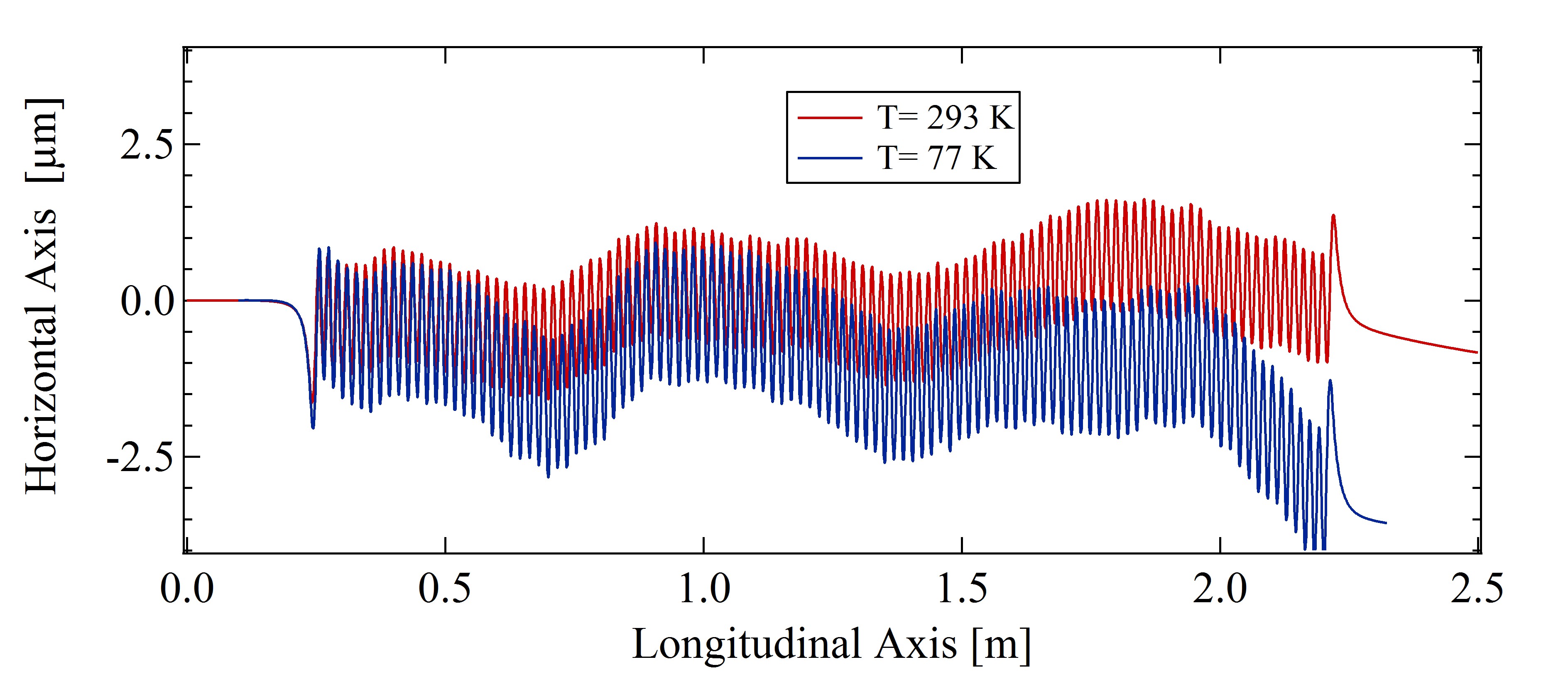}
    \caption{Electron beam trajectory calculated from the measured magnetic field. (Red curve) at room temperature 293 $K$, (Blue curve) at cryogenic temperature 77 $K$.}
    \label{Fig12}
\end{figure}

Figure \ref{Fig12} presents the electron trajectory at room temperature and cryogenic temperature calculated from the measured magnetic field. Despite the cooling down, the electrons trajectory position is kept below 4 $\mu m$ along the undulator.

The cryogenic undulator RMS phase error has been corrected (using shims) at room temperature to 2.8$^o$. After the cooling down, the RMS phase error is increased to 9.1$^o$ because of mechanical contractions. The rods are contracted vertically by 1 $mm$, the variation of this contraction between rods causes in-vacuum girder deformation and then phase error degradation. Mechanical shims have been used to modify the vertical position of the 24 rods in order to correct the RMS phase error and bring it down to 3 $^o$. The thickness of the shim is calculated using equation \ref{eqn7}.

\begin{equation}
\Delta B_z = B_0 \exp \left(\frac{\pi \Delta_g}{\lambda_u}\right)
\label{eqn7}    
\end{equation}

where $\Delta B_z$ is the magnetic field variation due to a gap variation of $\Delta_g$, and $B_0$ is the measured magnetic field before the gap modification.

\begin{figure}[!ht]
  	\centering
    \includegraphics[scale=.5]{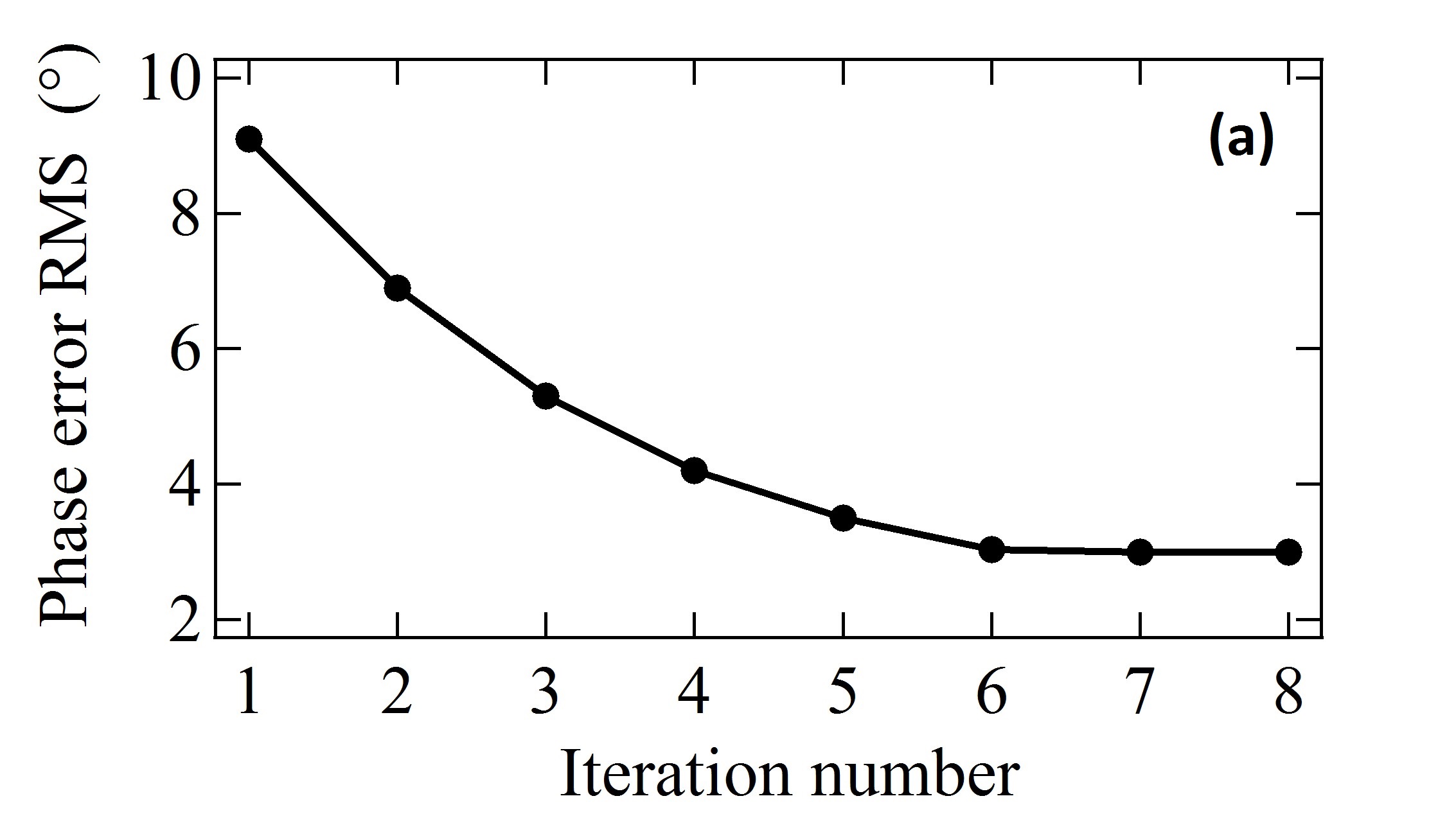}
    \includegraphics[scale=.5]{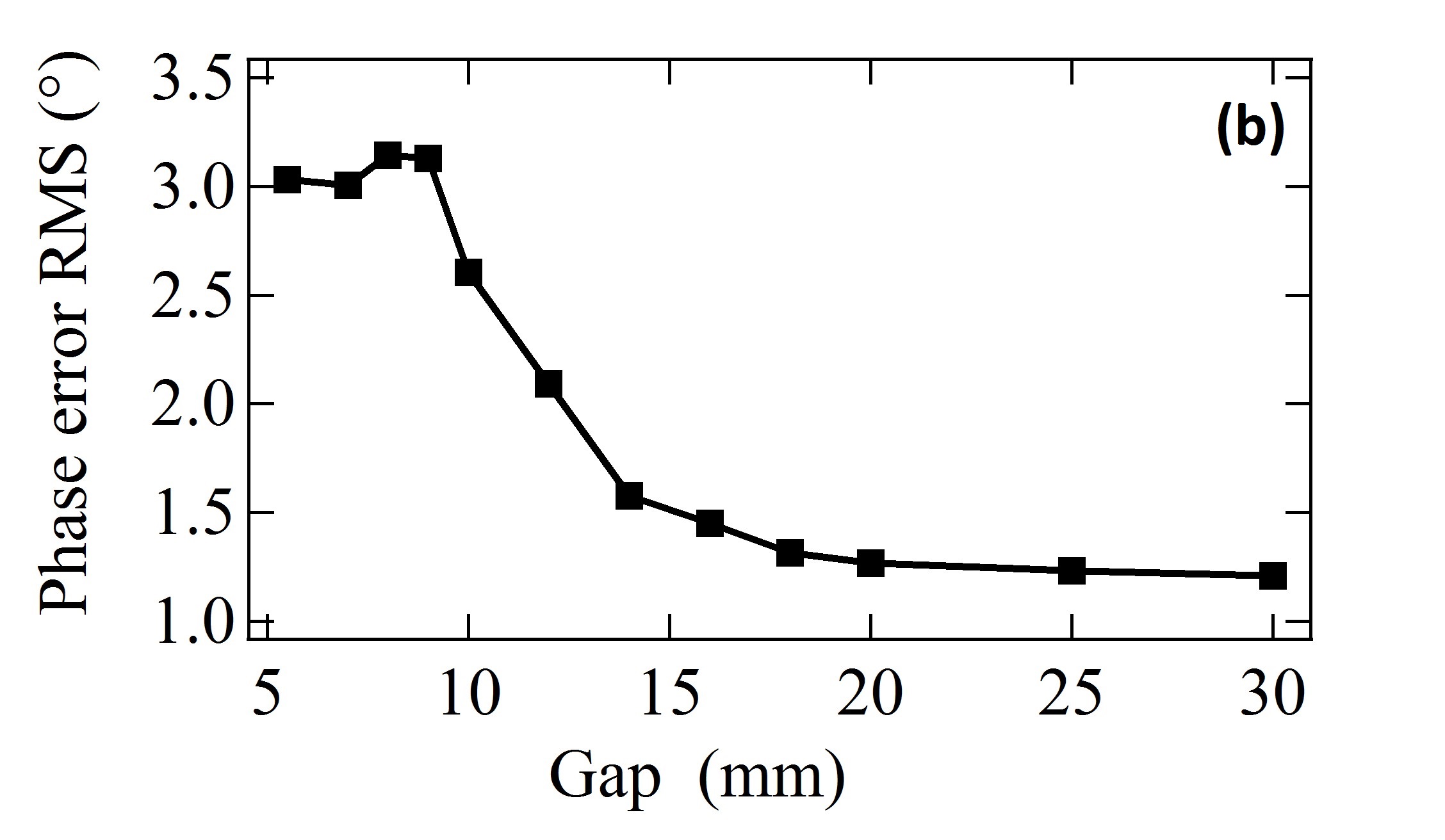}
    \caption{(a) Phase error corrections with rod vertical displacement at minimum gap of 5.5 mm. (b) Phase error variation versus gap after correction.}
    \label{Fig13}
\end{figure}

\newpage
Figure \ref{Fig13} (a) presents the improvement of the phase error calculated from the magnetic field measurement at 77 K during rods shimming iterations. The 3$^o$ RMS phase error is obtained from the $6^{th}$ iteration; the $7^{th}$ and $8^{th}$ iterations do not improve further the phase error.\\
Figure \ref{Fig13} (b) presents the phase error variation versus gap. The value at minimum gap of 5.5 mm is 3$^o$ RMS, the maximum value is a 3.2$^o$ RMS at gap 8 mm.

\begin{figure}[!ht]
  	\centering
    \includegraphics[scale=.4]{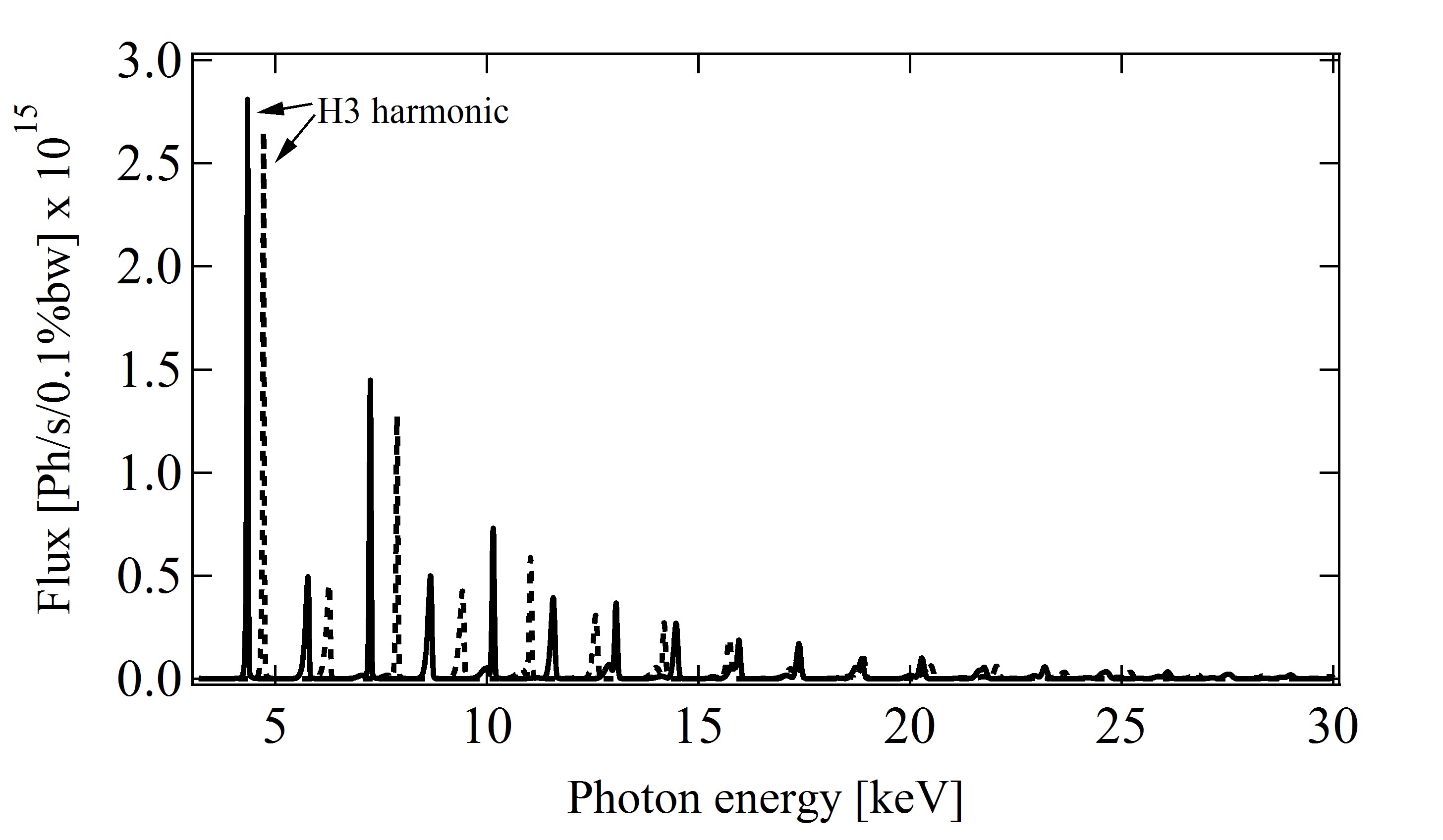}
    \caption{U18 spectrum calculated with SRW code at a distance of 17 m from the source point through a slit of 0.1 x 0.1 $mm^2$ at minimum gap of 5.5 $mm$. (Dash line) at room temperature of 293 $K$, (Solid line) at cryogenic temperature of 77 $K$. Beam characteristics shown in table \ref{Table12}.}
    \label{Fig14}
\end{figure}

Figure \ref{Fig14} presents the spectrum of the cryogenic undulator at room temperature and cryogenic temperature of 77 $K$ calculated from the measured magnetic field at minimum gap. At 77 $K$ the photon flux increases by 24$\%$ for the seventh harmonic and 34$\%$ for the ninth harmonic.

\section{\label{sec:level6}Operation of the cryogenic system on the SOLEIL storage ring}

\subsection{\label{subsec:level61}Commissionning with the electron beam}
\begin{figure}[!ht]
  	\centering
    \includegraphics[scale=.81]{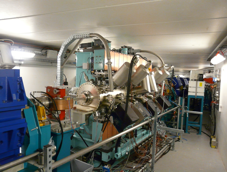}
    \caption{The cryogenic undulator installed in the SOLEIL storage ring straight section.}
    \label{Fig15}
\end{figure}

Figure \ref{Fig15} presents the cryogenic undulator installed in the long straight section SDL13 in SOLEIL storage ring. This straight section has been modified to allow for the installation of two canted in-vacuum undulators in order to extract two different long beam lines NANOSCOPIUM and ANATOMIX  \cite{LoulergueIPAC2009}.

\subsubsection{\label{subsec:level61}Alignment of the undulator with the electron beam}

\begin{figure}[!ht]
\centering
\includegraphics[scale=0.6]{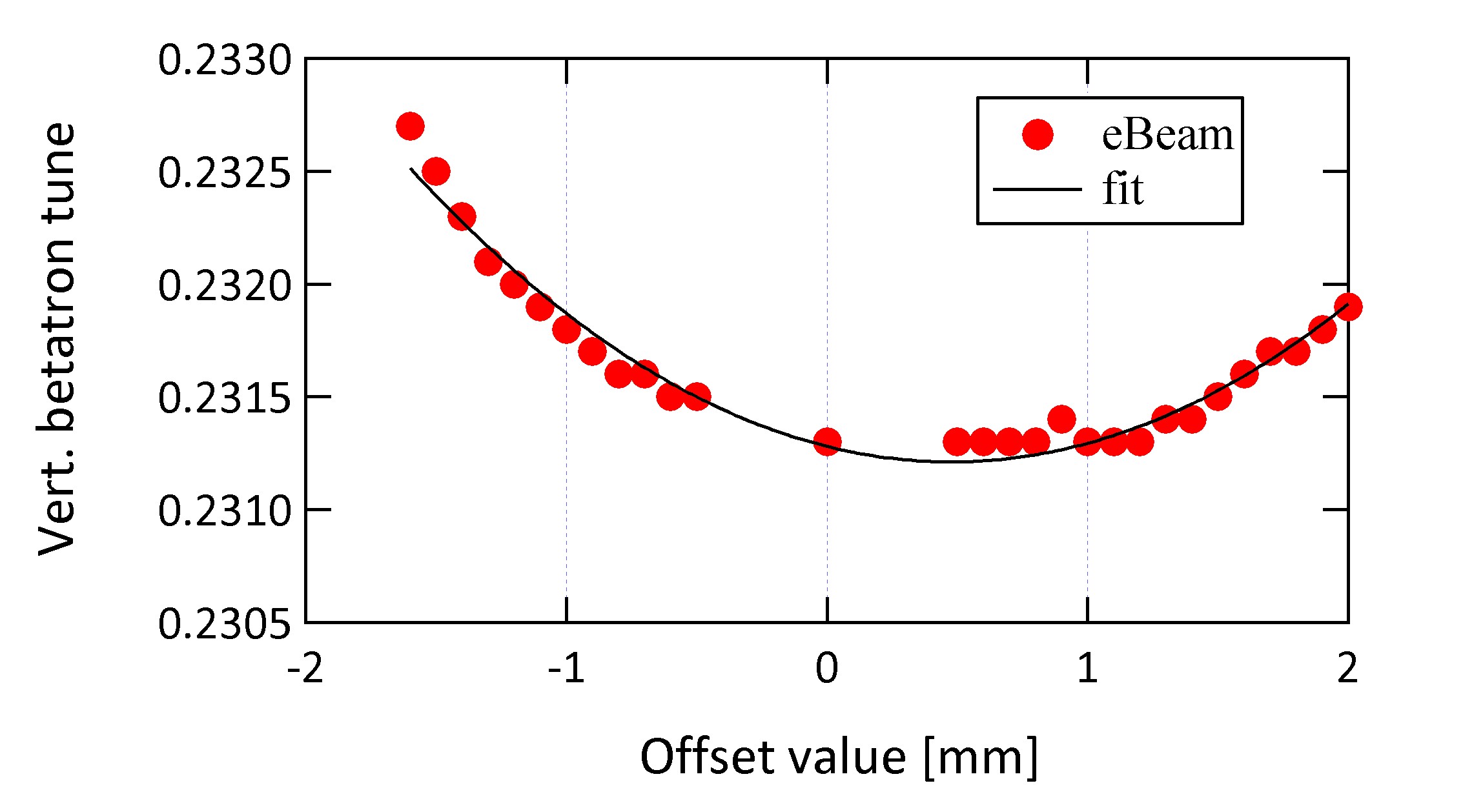}
\caption{Vertical betatron tune versus offset}
\label{Fig:AligUwE}
\end{figure}

After the installation and alignment of the undulator in the tunnel, the magnetic axis has to match with the electron beam path in the straight section. A vertical misalignment of the undulator has a very strong effect on the magnetic field and leads to a non-optimized phase error and trajectories. To perform this optimization with the electron beam, the whole carriage is moved in the vertical direction (offset movement) between +/-2 $mm$ at a fixed gap of 5.5 $mm$ and for each offset value, the vertical tune values are recorded (see Fig. \ref{Fig:AligUwE}. The magnetic axis corresponds to the offset value where the magnetic field value is miniumum so as the vertical tune ($B_0^2$ scaled). The reproducibility of this measurement is 50 $\mu m$.

The magnetic axis can also be determined by measuring the decay of the electron beam current versus offset.

\subsubsection{\label{subsec:level61}Vacuum and thermal evolutions}

\begin{figure}[!ht]
  	\centering
    \includegraphics[scale=.5]{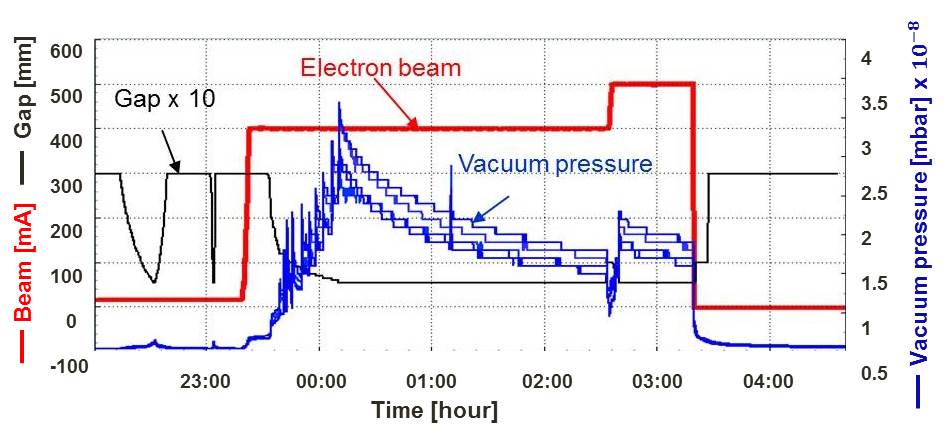}
    \caption{Undulator vacuum pressure variation versus gap at high electron beam current of 400 $mA$.}
    \label{Fig16}
\end{figure}

Figure \ref{Fig16} presents the evolution of the undulator vacuum pressure during the first closing of the gap at high beam current of 400 $mA$. The gap has been closed in only 40 min and the maximum pressure reached  3.10$^{-8}$ $mbar$.

\begin{figure}[!ht]
  	\centering
    \includegraphics[scale=.55]{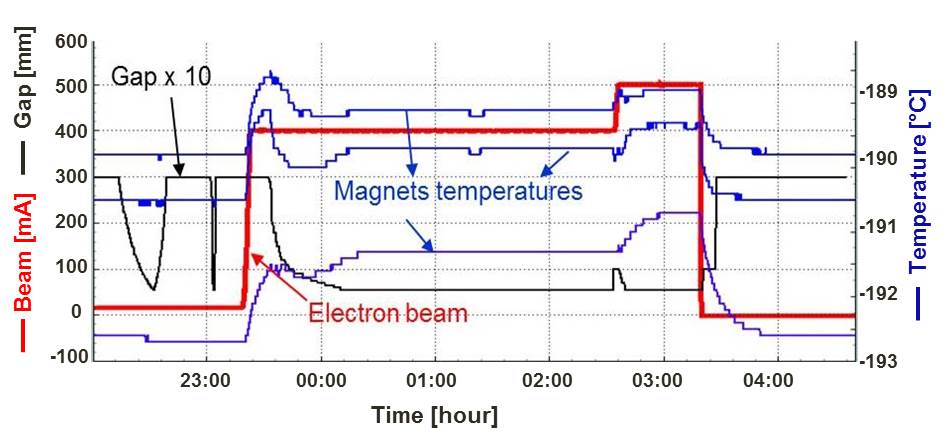}
    \caption{Permanent magnet temperature variation versus gap at high electron beam current of 400 $mA$.}
    \label{Fig17}
\end{figure}

Figure \ref{Fig17} displays the evolution of the undulator permanent magnet temperature during the first closing at the minimum gap at high beam current of 400 $mA$. The average permanent magnet temperature without the electron beam is 82 $K$. It increases only by 2 $K$ at high electron beam of 400 $mA$. The temperature dispersion between the permanent magnets is less than 3 $K$, and has a negligible effect on the magnetic field variation value.

\subsection{\label{subsec:level62}Effects on the electron beam}
\begin{figure}[!ht]
  	\centering
    \includegraphics[scale=.4]{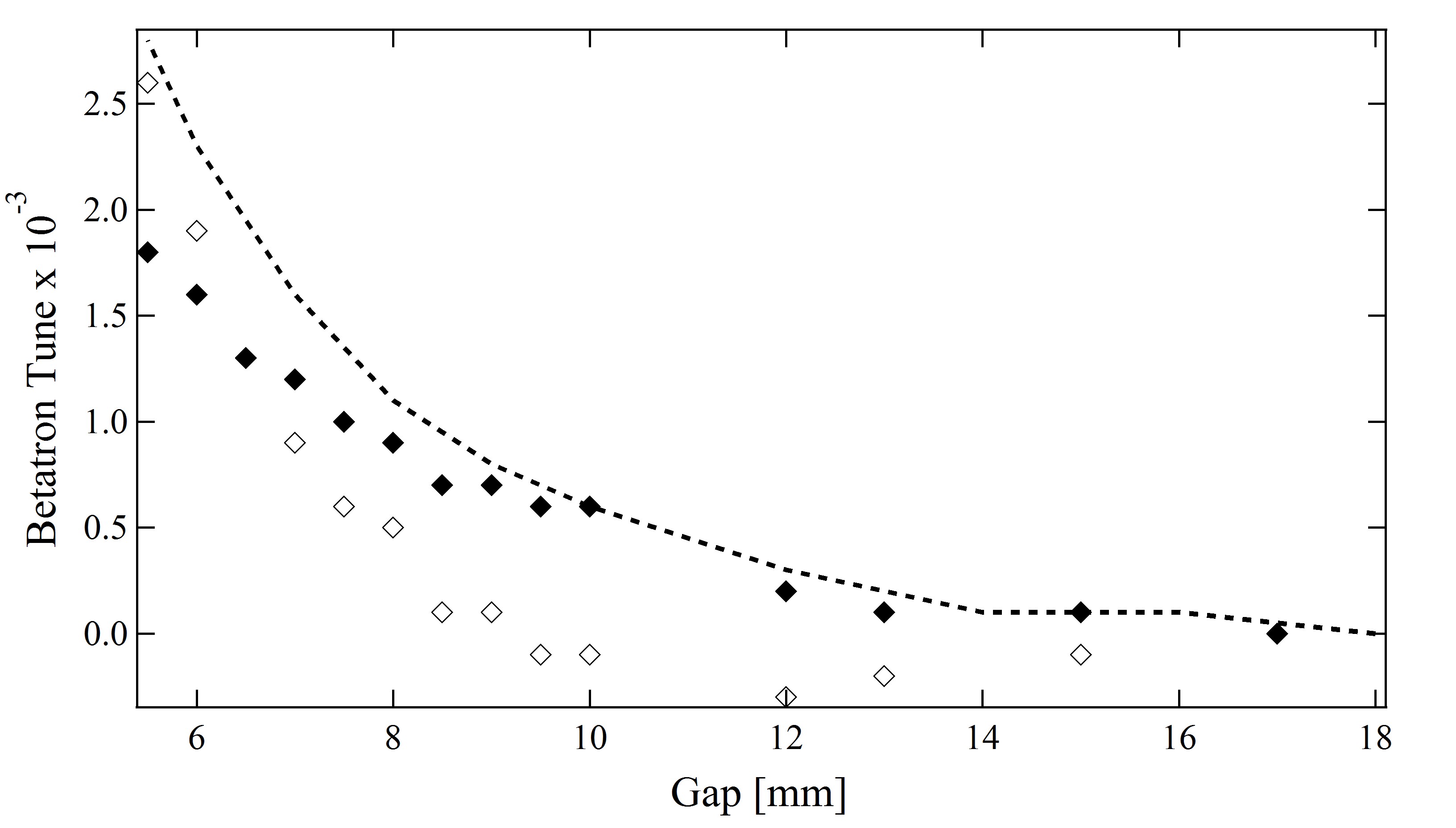}
    \caption{Variation of the measured horizontal (white diamonds) and vertical (black squares) tune variation as a function of gap. 
The vertical tune variation is compared to the expected one from measured vertical field scaling (dashed line).}
    \label{Fig19}
\end{figure}

The focusing effect of the cryogenic undulator on the electron beam has been measured as a function of gap. The vertical focusing is generated on one hand by the vertical field ($B_0^2$ scaling) and on the other hand by the normal integrated gradient measured on the magnetic bench. The horizontal focusing is generated only by the normal integrated gradient measured on the magnetic bench. Figure \ref{Fig19} presents the variation of both horizontal and vertical tunes versus gap. As expected from magnetic measurements, a significant horizontal focusing is observed due to the non-zero normal integrated gradient measured on the bench (+200 $G$). Actually the magnetic correction strategy was to perfectly cancel the skew integrated gradient term to the detriment of the normal one. The vertical focusing is then reduced compared to the one expected from the measured vertical field scaling.

\subsection{\label{subsec:level63}Spectral performance on the beamline}

The U18 cryogenic undulator is in use by the NANOSCOPIUM long beamline. This 155 $m$ long multimodal beamline  \cite{somogyi2011scanning, somogyi2013status} is dedicated to scanning hard X-ray nano-probe experiments in the 30 $nm$ $-1$ $\mu m$ spatial resolution range by combining X-ray fluorescence (XRF), absorption spectroscopy (XAS), and phase-contrast imaging). NANOSCOPIUM aims at reaching 30-200 $nm$ resolution in the 5-20 $keV$ energy range for routine user experiments. The beamline design tackles the tight stability requirements of such a scanning nano-probe by creating an overfilled secondary source, implementing all horizontally reflecting main beamline optics, and constructing a dedicated high stability building envelope. This beamline provides high sensitivity elemental and sample morphology mapping with down to 30 $nm$ spatial resolution by fast scanning spectro-microscopy combined with absorption, differential phase contrast and dark field imaging and electron density mapping by coherent imaging techniques. The typical scientific fields cover biology and life sciences, earth- and environmental sciences, geo-biology and bio-nanotechnology. The beamline is especially well suited for studies seeking information about highly heterogeneous systems at multiple length scales also in natural or in operando conditions.

\subsubsection{\label{subsec:level63}Measured undulator spectrum}

\begin{figure}[!ht]
  	\centering
    \includegraphics[scale=.4]{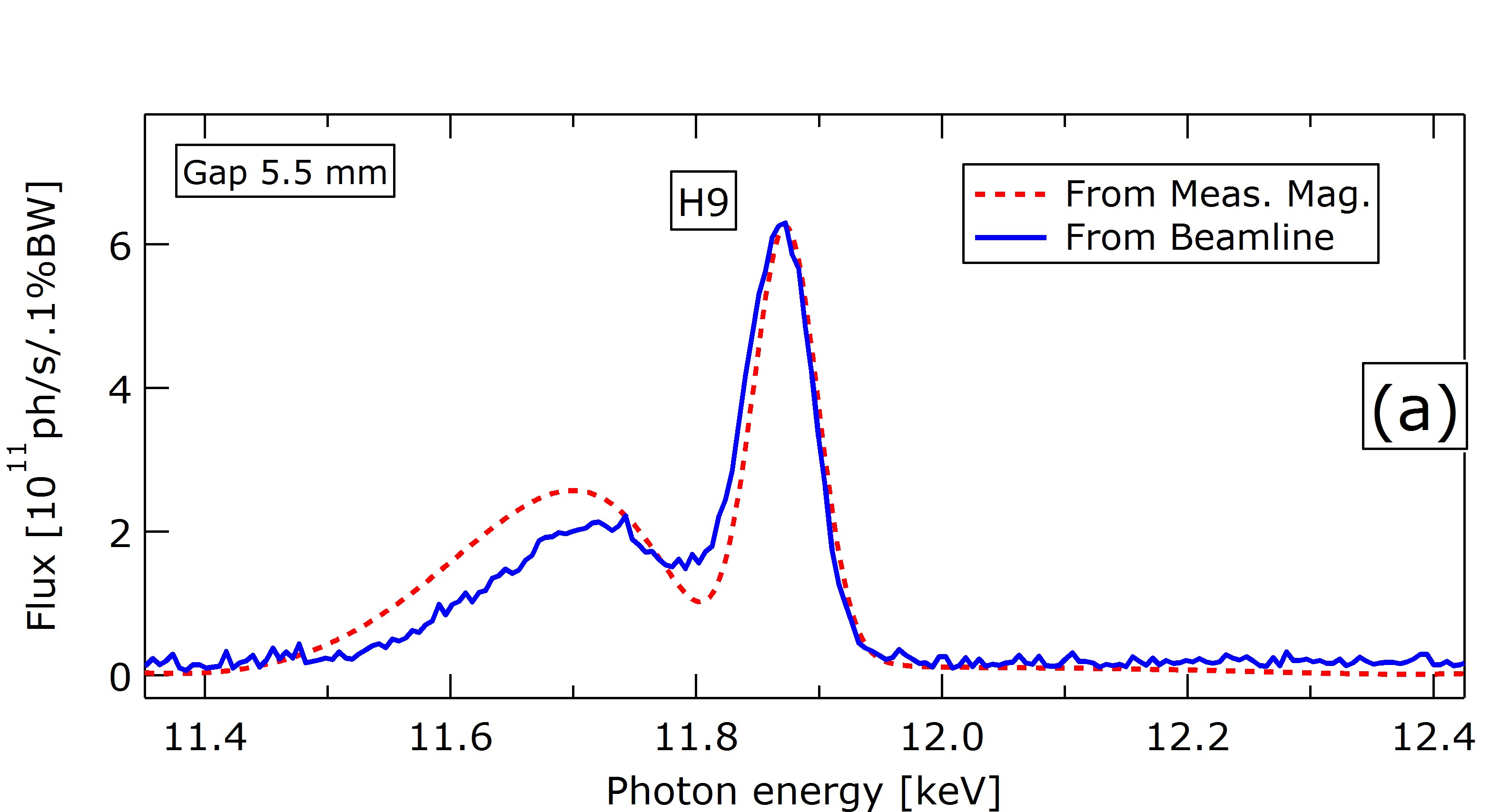}
    \includegraphics[scale=.4]{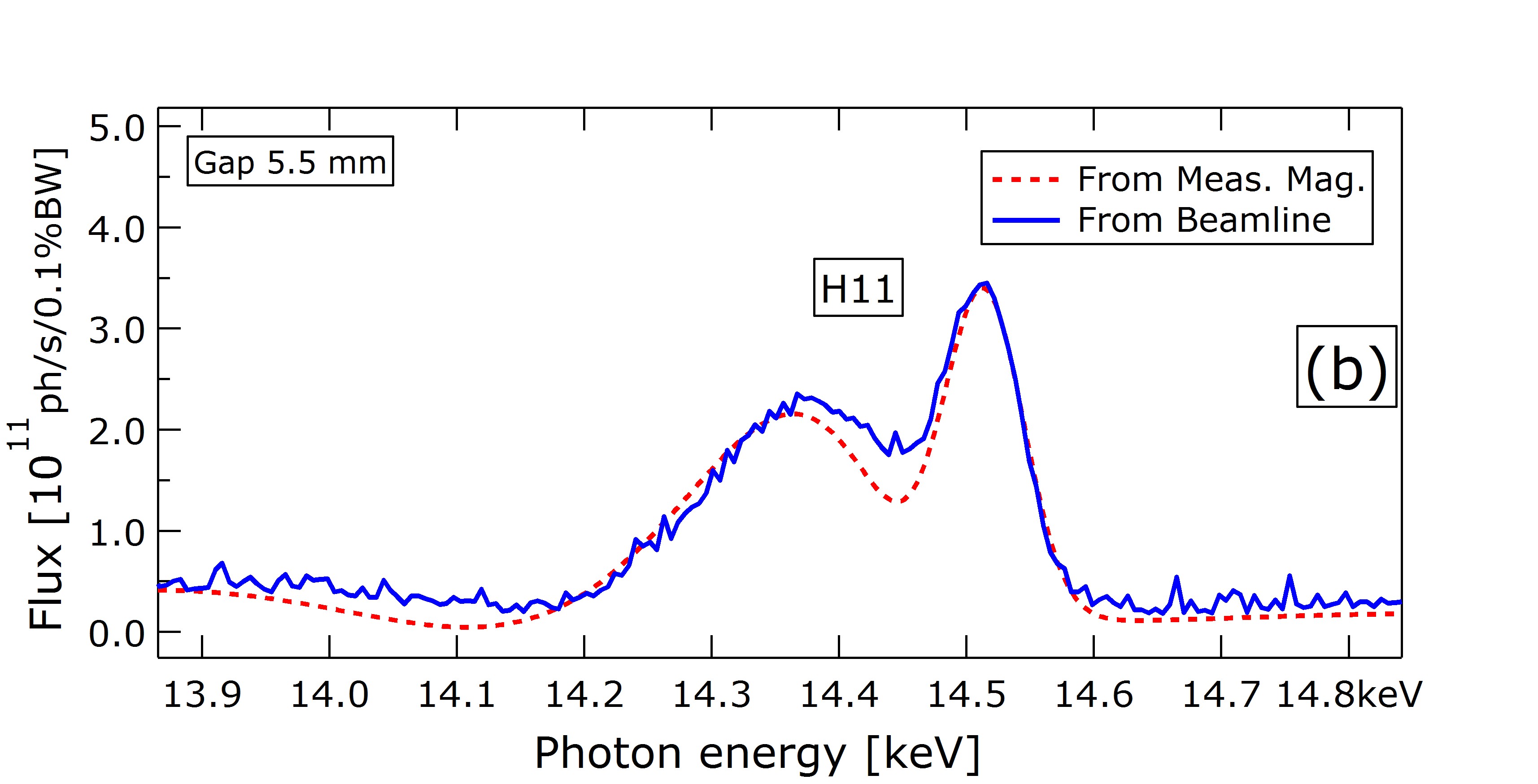}
    \includegraphics[scale=.4]{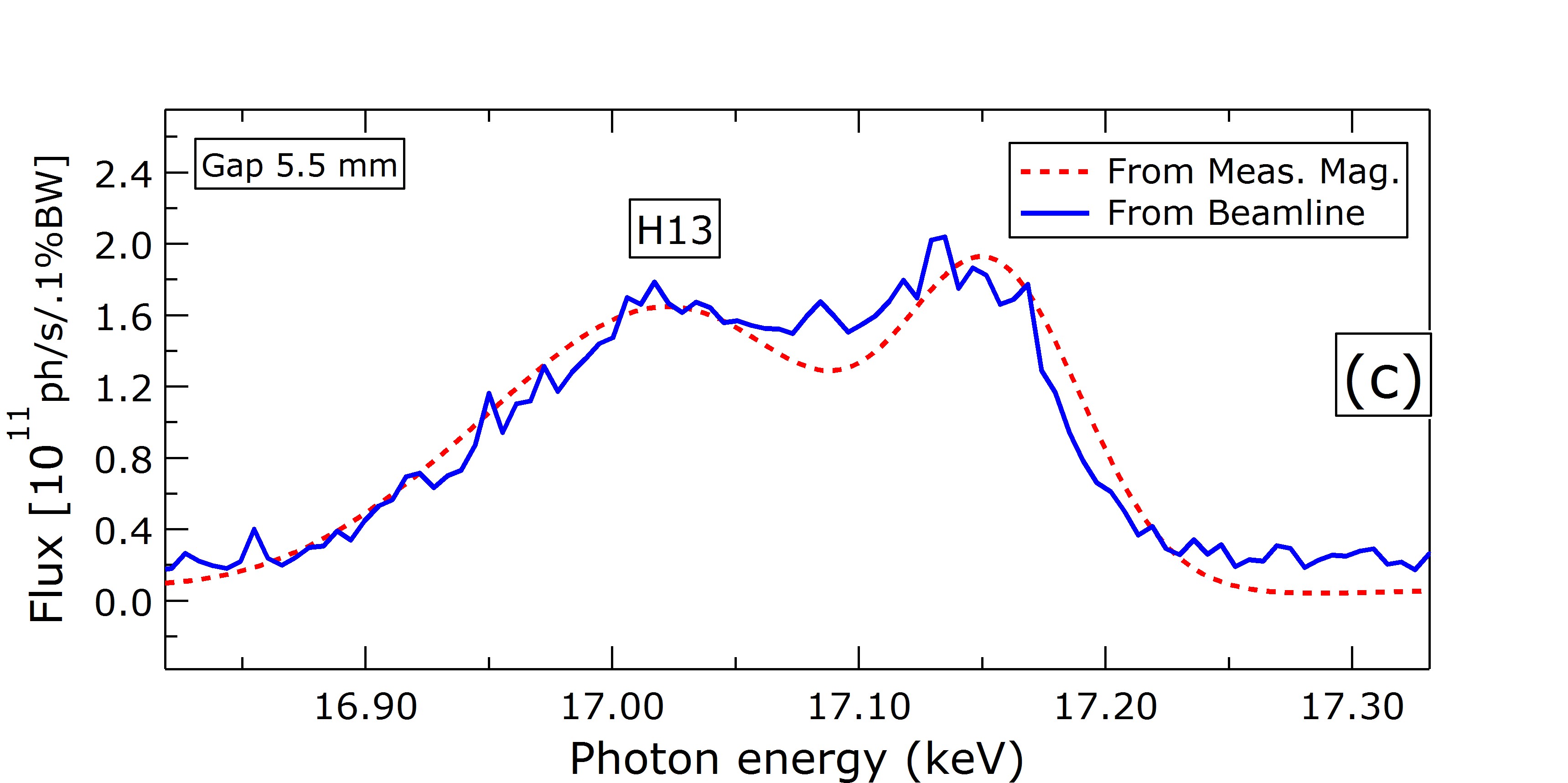}
    \caption{Spectra measured on the beamline and calculated from magnetic measurements at 5.5 $mm$ gap through a 0.1 $mm$ x 0.1 $mm$ aperture at a distance of 77 $m$ from the undulator. Electron beam parameters are shown in table \ref{Table12}, with horizontal emittance = 4.6 $nm$, vertical emittance = 0.0473 $nm$, $\beta_x$= 7.579 $m$, $\beta_z$= 2.34 $m$, $\alpha_x$= -0.851 $rad$, $\alpha_z$= $-0.033$. (a): $9^{th}$ harmonic, (b): $11^{th}$, and (c): $13^{th}$ harmonic.}
    \label{Fig18}
\end{figure}

The spectrum emitted by the cryogenic undulator U18 has been  measured on the NANOSCOPIUM long beamline. Figure \ref{Fig18} shows the photon flux on the harmonic H9 at 11.873 $keV$ of the spectrum measured on the beamline and compared to the one calculated from the magnetic measurements using the parameters of table \ref{Table12}.

A very good agreement has been found in terms of bandwidth between the measured spectrum on the beamline and the calculated one from the magnetic measurements. This result confirms the small optimized phase error of 3 $^o$ RMS obtained thanks to the in-situ corrections without dismounting of the vacuum chamber.

\subsubsection{\label{subsec:level63}Photon Beam Based Alignment of the undulator}

After the installation of an undulator in a storage ring, the magnetic axis is well aligned with the electron beam, to ensure the best performance possible. When the user beamline starts to be commissioned, the undulator alignment can be performed in a more precise manner in looking at the undulator harmonic shapes for different electron translation and angle bumps.

After some time, misalignment might occur and it is worth checking with the photon beam alignment. Thus experiments have been done on the spectrum of the undulator radiation with the NANOSCOPIUM beamline at SOLEIL, to ensure good alignment.


\subsubsection{Offset Optimization}
One illustrates here an offset optimization while monitoring 
the spectrum of the undulator radiation on the NANOSCOPIUM long section beamline with a window aperture (0.2 $mm$ $\times$ 0.8 $mm$) placed 77 $m$ away from the undulator,  and a photodiode placed at a distance of 83 $m$. The adjustment has been done by monitoring the $11^{th}$ harmonic, since high harmonics are very sensitive to any change in the beam parameters or the undulator characteristics.
The offset was varied, as in moving the girders up or down, while keeping the magnetic gap constant (5.5 $mm$).

\begin{figure}[!ht]
\centering
\includegraphics[scale=0.4]{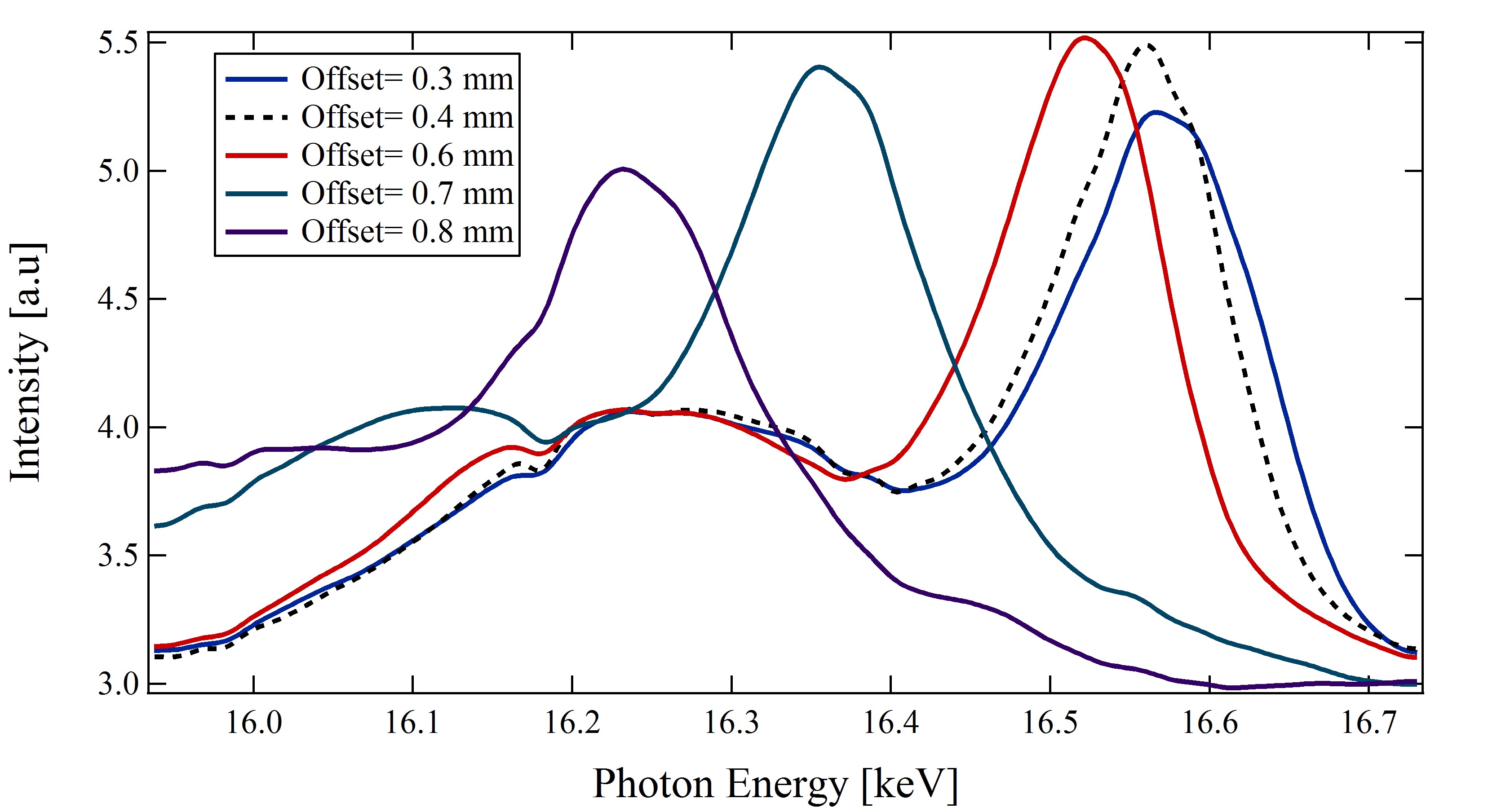}
\caption{Spectra measured (uncalibrated energy monochromator) on the NANOSCOPIUM beamline through a window aperture of 0.2 x 0.8 mm$^2$ placed at a distance of 77 $m$ from the undulator with electron beam parameters shown in Table \ref{Table12}, but with current I= 17 $mA$, $\beta_x$= 7.507 $m$, $\beta_z$= 2.343 $m$, $\alpha_x$= -0.846 $rad$, $\alpha_z$= 0.03 $rad$.}
\label{Fig:11harmonic}
\end{figure}

\begin{figure}[ht]
\centering
\includegraphics[scale=0.4]{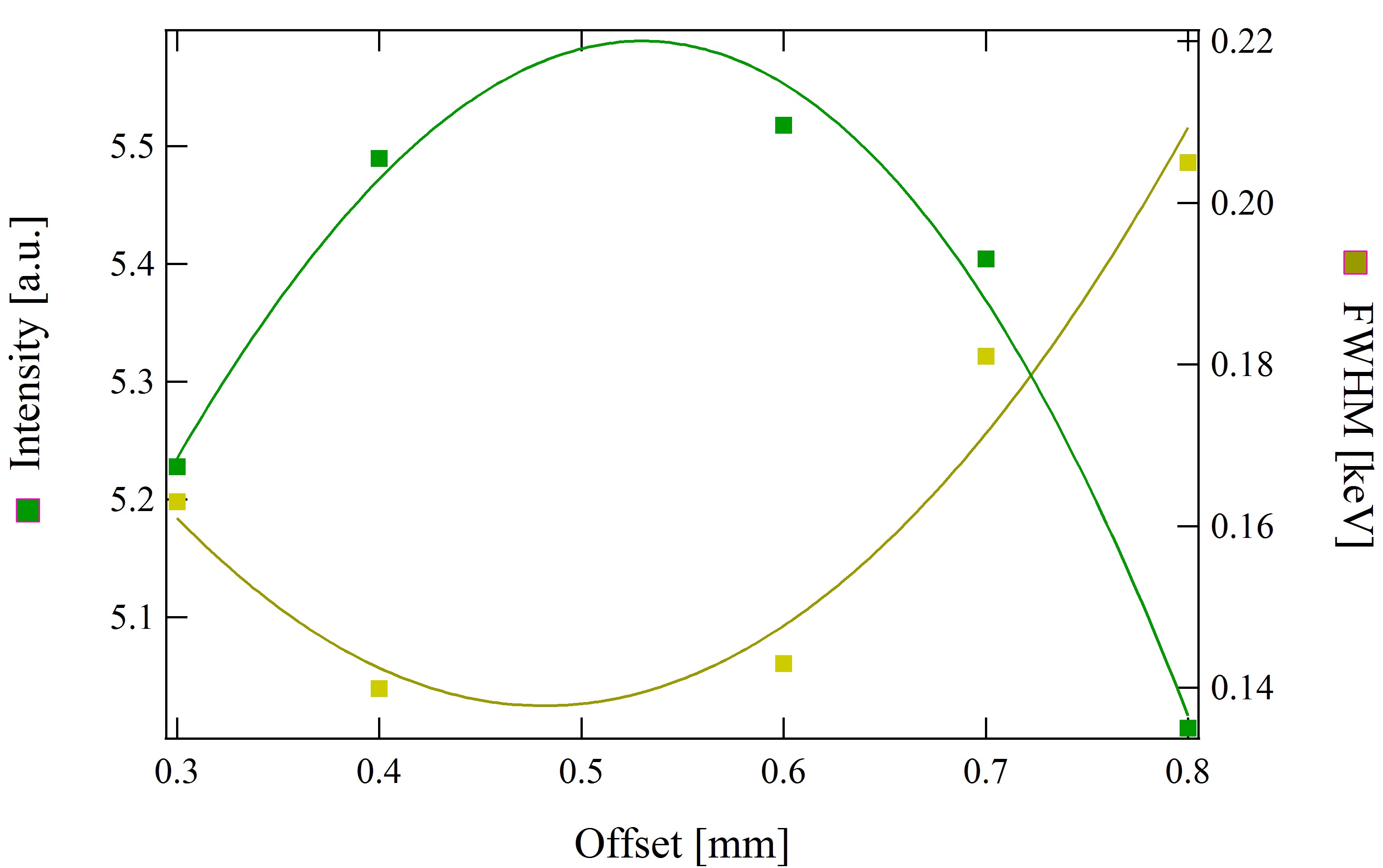}
\caption{Intensity and Full Width Half Maximum (FWHM) of the $11^{th}$ harmonic as a function of the offset that varied from 0.3 $mm$ up to 0.8 $mm$.}
\label{fig:OffsetM}
\end{figure}

Fig \ref{fig:OffsetM} shows the evolution of the intensity and bandwidth versus the offset variation. The best alignment is found when the highest intensity with the lowest half band width are observed. Originally the offset was at 400 $\mu m$ and is considered as the reference case for the new calculations shown in Fig \ref{fig:OffsetM}, and the new one is found to be at $\sim$500 $\mu m$. Therefore U18 offset had been adjusted to this value to achieve the best alignment possible as in highest intensity. 

These photon beam based alignment techniques are also carried out with electon (translation, angle) in the undulator, for a precise knowledge of the magnetic axis.

\subsubsection{Taper Optimization}
\begin{figure}[ht]
\centering
\includegraphics[scale=0.4]{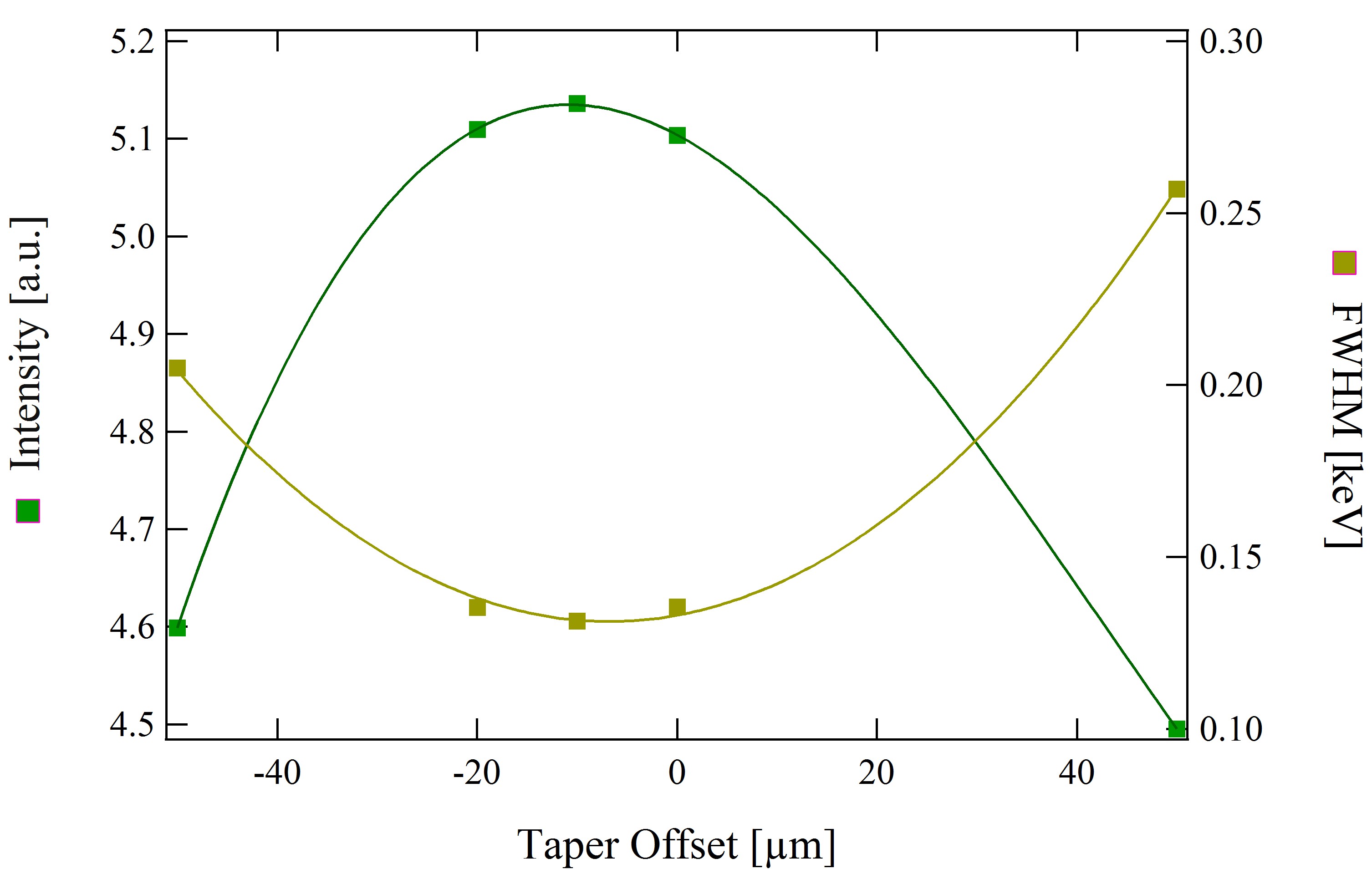}
\caption{Intensity and Full Width Half Maximum (FWHM) of the $11^{th}$ harmonic as a function of the taper offset.}
\label{Fig:Taper}
\end{figure}

Undulator tapering technique (i.e. variation of the peak field along the undulator axis) is vastly used in the FEL to improve its output power, by keeping the resonant radiation wavelength constant, despite the deviation in the energy of the electron beam. As for Synchrotron Radiation, it is one solution to perform an energy scan with a more constant intensity. One way to modify the field is to slightly vary the gap at the entrance or exit of the undulator, either by closing or opening girders at the extremities. Fig \ref{Fig:Taper} shows the change of intensity of the $11^{th}$ harmonic as we change the taper value.

The highest intensity with the lowest band width is at -10 $\mu m$; i.e. the girders at the end of the undulator are closed by 10 $\mu m$. This optimization was useful for the beamline at a precision of $\sim$5 $\mu rad$ and increased the flux by $\sim 0.6 \%$.

\subsubsection{\label{subsec:level63}Five years of operation}

The undulator has been installed in the storage ring since August 2011. The cryosystem, which is installed out of the tunnel, is connected to the undulator thanks to 2 flexible lines of 15 $m$ each via a "valve panel". This panel was developed by SOLEIL enabling to change the cryosystem while keeping the undulator at cryogenic temperature. This operation has been done only once in July 2014 where an ice plug blocked the flow in a line between the panel and the cryosystem. After some tests, it appeared that this ice plug was in the panel which was bypassed until the end of the run. During the tests and the by-pass of the panel, the undulator started to warm-up leading to a beam loss of 4 hours due to a vacuum interlock. This failure was the only problem on this cryogenic device in 5 years which led to a beam loss. Parameters of the cryosystems have been adjusted with the beam so that the thermal longitudinal gradient remains below 3 $K$, while not decreasing the lifetime of the cryogenic pump. No disturbance have been observed on the beam due to the liquid flow, whatever the parameters used for the cryosystem. The undulator is warmed up one or twice a year, during electrical tests.

\section{\label{sec:level8}Conclusion}
A 2 m $Pr_2Fe_{14}B$ cryogenic undulator with 18 $mm$ period has been designed so that it suits for the NANOSCOPIUM long section beamline of SOLEIL. The use of $Pr_2Fe_{14}B$ permits to operate at the liquid nitrogen temperature without undergoing SRT effects resulting in an increase of the magnetic field peak, and attain a good thermal stability along the undulator. In contrast, $Nd_2Fe_{14}B$ based CPMU requires thermal resistances distributed along the girders for an operation at 140 $K$. The CMPU has been built, optimized at room temperature. It has been further corrected at cryogenic temperature using an additional in-vacuum measurement bench leading to a rms phase error of $3^o$. 

After the installation at SOLEIL storage ring, it has been commissioned for different electron beam filling modes and different gaps. The thermal behavior of the undulator is stable because the maximum operation temperature variation for all electron beam filling modes and gap variations is less than 3 $K$ (instead of 26 $K$ for $Nd_2Fe_{14}B$). Effects on the electron beam are in agreement with the magnetic measurements.

The spectrum has been measured on the beamline and shows a very good agreement with the one calculated from the magnetic measurements. Photon beam based alignment for the NANOSCOPIUM beamline provides a very fine tuning of the undulator axis.
The spectra are also very nice tools for the taper optimization allowing for a precision of 5 $\mu rad$.

\section*{\label{sec:level9}Acknowledgments}
We would like to thank Joel Chavanne from ESRF for his kind support, and members of the Accelerator and Engineering Division of SOLEIL led by A. Nadji. The authors are also very grateful for the support of J. M. Filhol, the former head of the Accelerator Division of SOLEIL and the European Research Council for the advance grant COXINEL (340015).

\end{document}